\documentclass[final, 3p, times]{elsarticle}
\usepackage{lineno,hyperref}
\usepackage{amssymb}
\usepackage{amsmath}
\usepackage{epsfig}
\usepackage{graphicx}
\usepackage{psfrag}
\usepackage{pstricks}
\usepackage{algorithm}
\usepackage{subfigure}
\usepackage{pst-grad}
\usepackage{pst-node}
\usepackage{pstcol,multido}
\usepackage{float}
\usepackage{algorithm}
\usepackage{algorithmic}
\usepackage{multirow}
\usepackage{longtable}
\usepackage{epstopdf}
\makeatletter

\newcommand{\Rmnum}[1]{\expandafter\@slowromancap\romannumeral #1@}
\newtheorem{remark}{Remark}[section]

\makeatother
\newtheorem{theorem}{{Theorem}}

\newtheorem{proof}{Proof}

\journal{}

\begin{document}
\begin{frontmatter}

\title{Relaxed Total Generalized Variation Regularized Piecewise Smooth Mumford-Shah Model for Triangulated Surface Segmentation}

\author[address1]{Huayan Zhang}
\author[address1]{Shanqiang Wang}
\author[address2]{Xiaochao Wang}
\address[address1]{School of Computer Science and Technology, Tiangong University, Tianjin, China}
\address[address2]{School of Mathmatics, Tiangong University,Tianjin, China}

\begin{abstract}
The Mumford-Shah (MS) model is an important technique for mesh segmentation. Many existing  researches focus on piecewise constant MS mesh segmentation model  with total variation regularization, which pursue the shortest length of
   boundaries. Different from previous efforts, in this article, we propose a novel piecewise smooth MS mesh segmentation model by utilizing the relaxed total generalized variation  regularization (rTGV). The new model assumes that the feature
   function of a mesh can be approximated by the sum of  piecewise constant function and asmooth function, and the rTGV regularization is able
  to  characterize the high order discontinuity of the geometric structure. The
   newly introduced  method is effective in segmenting meshes with irregular structures and getting
   the better boundaries rather than the shortest boundaries. We solve the new model by  alternating
   minimization and alternating direction method of multipliers (ADMM).  Our algorithm is discussed from several aspects, and comparisons with
   several state-of-art methods. Experimental results show that our method can yield competitive results when  compared to other approaches. In addition,  our results compare
   favorably to those of the several state-of-art techniques when evaluated on the Princeton Segmentation Benchmark.
   Furthermore, the quantitative errors and computational costs confirm the robustness and efficiency of the proposed
   method.
\end{abstract}


\end{frontmatter}




\section{Introduction}
Mesh segmentation is a fundamental task  in geometric modeling and computer graphics community. It has many applications
such as parameterization, simplification, shape retrieval, multiresolution modeling, skeleton extraction and so on.

For a given mesh $\mathrm{M}$ with region $\Omega$,  mesh segmentation aims to  decompose the mesh region
$\Omega$ into $\mathbf{K}$ disjoint connected piecewise smooth subsets $\Omega_k$ with a union of
smooth boundaries $\Gamma$ such that
$$\Omega = \Omega_1\cup\ldots\cup \Omega_\mathrm{\mathbf{K}}\cup\Gamma, \ \ \ \  \Omega_i\cap\Omega_j = \emptyset.$$
The pioneered Mumford-Shah (MS)  segmentation model \cite{Mumford89} finds an optimal piecewise smooth approximation
$u$ of $f$ by the following minimization problem:
\begin{equation}\label{MS}
\min_{\mathbf{u},\Gamma} |\Gamma| + \beta\int_{\Omega\setminus\Gamma}|\nabla u|^2dx + \alpha\int_{\Omega}|f-u|^2dx,
\end{equation}
where $f$ is the input feature function of $\mathrm{M}$, $u$ is a piecewise smooth approximation of $f$,
such that $u$ varies smoothly within each $\Omega_i$ and discontinuously across the boundaries of
$\Omega_i$, and $|\Gamma| $ stands for the total boundary length.

In practice, it is very challenging to solve \eqref{MS} due to its non-convex and the discontinuous of the integral
regions.  A classical reduced version of model \eqref{MS} can be obtained by restricting $u$ into a piecewise
constant function, i.e. $u = \sum\limits_{k=1}^{K}\mu_k\gamma_{k}$, $\mu_k$ and $\gamma_{k}$ are the
feature value and  the indicator function of region $\Omega_i$, respectively. The reduced piecewise constant
MS model can be formulated as follows
\begin{equation}\label{PCMS}
\min_{\{\mu_i\},\{\Omega_i\}} \sum_i|\partial\Omega_i| +  \alpha\sum_{i=1}^{\mathbf{K}}\int_{\Omega}|f-\mu_i|^2\gamma_{i}dx.
\end{equation}
In addition, \cite{Lellmann09,Lellmann11} further reformulated \eqref{PCMS}  into the following optimization problem by
introducing continuous multi-class label $\mathbf{u}=\{u_k\}$, with $u_k\in [0,1]$ and $\sum\limits_k u_k = 1.$
\begin{equation}\label{MC_MS}
\min_{\{\mu_i\},\{\mathbf{u}\}} \int_{\mathrm{\Omega}}|\nabla\mathbf{u}|dx +  \alpha\sum_{i=1}^{\mathbf{K}}\int_{\Omega}|f-\mu_i|^2u_idx,
\end{equation}
where the first term is based on the fact the perimeter of $\{\Omega_i\}$ can be expressed as the total
variation of the indicator function \cite{chan2006algorithms}.

The piecewise constant MS model \eqref{PCMS} and
\eqref{MC_MS} have been successfully applied in image segmentation
\cite{Chambolle95,Chan02,Lie06,lie2006variant,Brown12} and mesh segmentation \cite{Zhang12,zhang2018new}.
However, for images with intensity inhomogeneity and surfaces with irregular regions, the piecewise constant MS
model may  get into local minima. Many variants piecewise smooth MS models are proposed to process images with intensity inhomogeneity
\cite{Vese2007,gu2016generalizing,li2010variational,gu2013efficient,jung2017piecewise,li2020tv}.
By treating the intensity inhomogeneity as small perturbation of piecewise constant intensities, the following model is
proposed by decomposing the  piecewise smooth function  into a piecewise constant function and a smooth function.
\begin{equation}\label{ps_function}
f(x) = \sum\limits_{k} b(x)\mu_k + n(x),  x\in\Omega,
\end{equation}
where $b(x)$ is a smooth function, and $n(x)$ is the noise.

Based on \eqref{ps_function},
\cite{li2010variational} reformulated the MS model \eqref{MS} as the following minimization problem:
\begin{equation}
\begin{split}
\min\limits_{\{b,\Omega_k,\mu_k\}}\sum\limits_{k}(|\partial\Omega_k| &+ \frac{\alpha}{2}\int_{\Omega_k}(f-b\mu_k)^2
+\omega(f-\mu_k)^2dx)\\
 &+\beta\int_{\Omega}|\nabla b|^2dx.
\end{split}
\end{equation}
In addition, the smooth function in \eqref{ps_function} can be defined as the following additive way
\cite{Vese2007,gu2013efficient,jung2017piecewise,li2020tv}
\begin{equation}\label{ps_function_add}
f(x) = \sum\limits_k\mu_k\gamma_k(x) + b(x) + n(x).
\end{equation}
Relaying on \eqref{ps_function_add}, \cite{Vese2007} proposed the following problem:
\begin{equation}
\begin{split}
\min\limits_{\{b,\Omega_k,\mu_k\}}\sum\limits_{k}(|\partial\Omega_k| + &\frac{\alpha}{2}\int_{\Omega_k}(f-\mu_k-b)^2dx)
+\frac{\beta}{2}\int_{\Omega}|\nabla b|^2dx \\
&+ \frac{\eta}{2}\int_{\Omega}|\nabla^2b|^2dx.
\end{split}
\end{equation}
\cite{jung2017piecewise} proposed a piecewise smooth MS model with the $L_1$ fidelity term instead of the $L_2$
term. \cite{li2020tv} built up the following piecewise smooth MS model by utilizing a non-convex
$L^p (0\leq p\leq 1)$ regularity term and a laplace smooth term.
\begin{equation}
\begin{split}
\min\limits_{\{b,\mathbf{u},\mu\}} \int_\Omega|\nabla\mathbf{u}|^pdx &+ \frac{\beta}{2}\int_{\Omega}|\triangle b|^2dx
 +\frac{\eta}{2}\int_{\Omega}|b|^2dx\\
 & + \alpha\sum_k\int_{\Omega_k}\langle (f-b-\mu_k)^2,\mathbf{u}\rangle dx.
\end{split}
\end{equation}

\begin{figure}[htbp]
\centering\includegraphics*[width=4.5in]{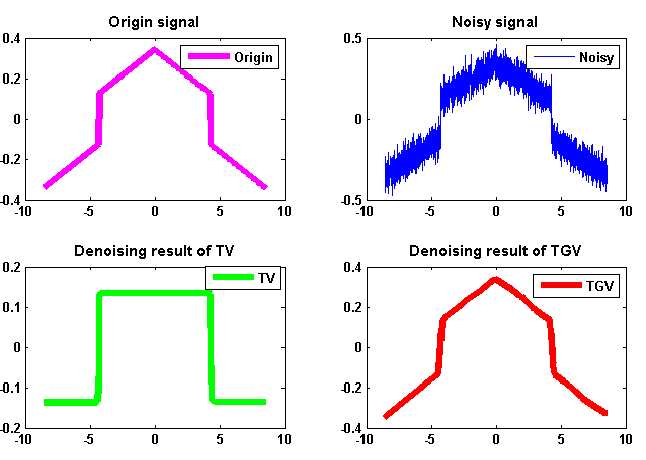}
\caption{{
The comparison between TV regularization and TGV regularization.
}
}\label{TV-TGV}
\end{figure}

The above piecewise smooth MS models have been demonstrated to be effective in dealing with images with intensity
inhomogeneity, which  try to
obtain the shortest length of boundaries $|\Gamma|$ with the total variation regularization (TV). However, as
the surface is irregular,  the shortest boundaries sometimes are not necessarily the optimal segmentation boundaries.  Recently, the second order relaxed total generalized variation (TGV/rTGV) regularization \cite{bredies2010total,zhang2022} on images and meshes has been proven to be  able to capture the first-order
discontinuity structures of a function and alleviate the stair-case effects of TV regularization effectively (See Figure~\ref{TV-TGV} for an example. For the one-dimensional noisy signal, the TV method cannot restore the vertex at the  first-order discontinuity, but TGV can preserve the vertex well)

Motivated by the good properties of the
TGV/rTGV regularization, in the paper, we consider the rTGV regularized piecewise smooth MS mesh
segmentation method to pursue better boundaries rather than just the shortest boundaries.
To the best of our knowledge, there is little research on rTGV regularized piecewise smooth MS model technique
for triangulated surface segmentation (\cite{TGVpointseg} proposes to use the piecewise linear TGV regularization for segmenting point cloud data. However, due to the lack of inherent topological structure in point clouds, the piecewise linear TGV regularization defined on them cannot guarantee convergence. As a result, the segmentation outcomes on point cloud data are often unsatisfactory) .  The variational model is a new discretization of the classical piecewise smooth Mumford-Shah model  by applying rTGV regularization  on triangle meshes. Besides, we use a simple alternating minimization and ADMM to solve the non-convex problem. Plenty of experiments showed this new segmentation method works well, which always gives satisfying segmentation results with correct human perception.

\section{Related work}
\subsection{Mesh segmentation}
Mesh segmentation is an important basic topic in geometric modeling and computer graphics community. With different applications, mesh segmentations can be divided into  surface-type and part-type segmentations.
  For surface-type segmentation, a mesh surface is partitioned into basic mathematical structures such as planes and spheres.
  For part-type segmentation, a mesh surface is partitioned into meaningful volumetric parts which respect human perception. So far a wide variety of algorithms have been developed for decomposing meshes. The interested reader can refer to excellent surveys
\cite{Luigi90,Shamir08,Agathos07,rodrigues2018part}, and a benchmark for evaluating mesh segmentation algorithms
\cite{Chen09}, as well as some recent developments, which  mainly include traditional variational-based method \cite{Oscar12,Zhang12,Kaick14,Nicolas18,tong2018spectral,zhang2020total} and learning-based segmentation \cite{Kalogerakis10,Kalogerakis17,le2017multi,lahav2020meshwalker,dong2023laplacian2mesh}.

 Isoline Cut \cite{Oscar12} proposes to  get the isolines of the concave-aware segmentation field with higher scores as the final contour.
The method is effective. However, it is not suitable for segmenting CAD meshes.
\cite{Zhang12} discretizes the piecewise constant Mumford-Shah image segmentation \cite{Mumford89}  to get a vertex-based mesh segmentation method.  As the Mumford-Shah model is a famous non-convex model, the solution severely depends on the initialization.
The segmentation method in \cite{Kaick14}  proposed a greedy algorithm by first over segmenting the models into approximate convex components, and then merging process and refining to achieve final segmentation. \cite{Nicolas18}proposed the Mumford-Shah mesh processing method by using the ambrosio tortorelli functional.  The authors of \cite{tong2018spectral} proposed a new spectral mesh segmentation method via $L_0$ minimization. The method adopt a coarse-to-fine strategy, first devising a  heuristic algorithm to
find a coarse segmentation, then proposing an optimization algorithm  to refine the segment boundaries. The algorithm can produce satisfying results, but it involves many parameters. \cite{zhang2020total} introduced a new total variation diffusion method to get the segmentation results, which is quite simple. However, the method is affected by initializations.  While obtaining good results, most variational-based segmentation methods involve  multiple step operations with higher computational cost. Some are still either unstable, or sensitive to initializations, or even need user interaction.

Recently, the deeping learning techniques have been studied for mesh segmentation. The pioneered learning-based segmentation method \cite{Kalogerakis10} proposed to get segmentation results using labeling techniques, which can get excellent results. The authors further improved the method by using the convolutional networks \cite{Kalogerakis17}.  The method in \cite{le2017multi} proposed  the coherent segmentation method  based on  the convolutional neural networks  and a two-layer long short term memory. In addition,  \cite{lahav2020meshwalker}  proposed  a novel representation of meshes  by random walks along
the surface, and then devised an end-to-end learning framework that realizes
the representation within Recurrent Neural Network.  Furthermore, by mapping the input mesh surface to the
multi-dimensional Laplacian-Beltrami space, \cite{dong2023laplacian2mesh} introduce the Laplacian2Mesh  framework for shape classification and segmentation. Overall, the learning-based methods can effectively process different types of data, and get better classification results. However, the learning-based methods rely on  large number of datasets, and the computational costs are higher.

In the paper,  we focus on the  rTGV regularized piecewise smooth MS mesh
segmentation method, which belongs to  the traditional variational-based method. The method combines the piecewise smooth MS model and rTGV regularization.
\subsection{Our contributions and paper organization}

 The existing researches on piecewise constant MS mesh segmentation method with TV regularization seek the shortest
length of boundaries. Different from previous work,  the goal of this paper is to investigate the piecewise smooth
Mumford-Shah  mesh segmentation model with the total generalized variation regularization, in order  to obtain the better
boundaries instead of the shortest boundaries.
\begin{itemize}
\item[-] We propose  the rTGV regularized piecewise smooth MS model for triangulated surface
segmentation method (containing the first order TV and the second order RTGV).
\item[-] The optimization problem is solved by the  alternating minimization and alternating direction
method of multipliers.
\item[-] Our algorithm is discussed and compared to  several state-of-art methods in various aspects.
Experimental results show that our piecewise smooth MS method can yield competitive results when  compared to
piecewise constant MS method, and other approaches on the Princeton Segmentation Benchmark. Furthermore, the quantitative errors and computational costs
confirm the robustness and efficiency of the proposed method.
\end{itemize}

The remainder of the paper is organized as follows. Section~2 gives some notations, differential operators on mesh and
the definition of TV/rTGV regularization. In section~3, we introduce our TGV regularized piecewise smooth MS mesh
segmentation method. Section~4 presents the details for solving our segmentation optimization problem.    In section~5, we present
the experiments and comparisons. Section~6 concludes the paper.

\section{The Relaxed Total Generalized Variation Regularization}

In the section, we introduce some notations followed by differential operators and the general total variation
regularization on triangulated surfaces.

\subsection{\label{notations}Notations}

Without loss of generality, we denote a triangulated surface with no degenerate triangles  as $M\subset\mathbb{R}^3$.
The set of
vertices, edges  and  triangles of $M$ are
denoted as $\{v_i:i=0,1,\cdots,\mathrm{V}_M-1\}$,
 and $\{e_i:i=0,1,\cdots,\mathrm{E}_M-1\}$, and $\{\tau_i:i=0,1,\cdots,\mathrm{T}_M-1\}$. Here
$\mathrm{V}_M$, $\mathrm{E}_M$  and $\mathrm{T}_M$ are the numbers of
vertices, edges and triangles, respectively.
If $v$ is an endpoint of an edge $e$, then we denote it as
$v\prec e$. Similarly, if $e$ is an edge of a triangle  $\tau$, it is denoted as $e\prec\tau $;
For each edge $e$, we choose a direction arbitrarily. If the direction of  $e$
is  consistent with its connecting triangle $\tau$, we denote $\mathrm{sgn}(\tau,e) = 1$, else $\mathrm{sgn}(\tau,e) = -1$.
Denote $N_{2}(\tau)$  as the second type $1$-neighborhood of triangle $\tau$, which are the triangles sharing some common
vertex with $\tau$.
\subsection{Differential operators on mesh}

We first present the definitions of gradient operator and divergence operator on the triangulated surface. Two types of
regularization are then given based on these operators (see \cite{Zhang15},
and \cite{zhang2022} for details).

We first denote the space $\mathbf{U}_M = \mathbf{R}^{\mathrm{T}_M\times n}$. For
$\mathbf{u} = (\mathbf{u}_0,\cdots,\mathbf{u}_{\tau},\cdots, \mathbf{u}_{\mathrm{T}_M-1}) \in \mathbf{U}_M$,
$\mathbf{u}_{\tau}=(u_{\tau,1},\cdots,u_{\tau,n})$ is a $n$-dimensional vector.
For each edge $e$, the gradient operators $\nabla$ restricted on each edge $e$ has the  following form:
\begin{eqnarray}\label{grad_M} (\nabla\mathbf{u})_e = \left\{\begin{array}{ll}
 \sum\limits_{\tau\prec e}\mathbf{u}_{\tau}\mathrm{sgn}(\tau,e),  & e\nsubseteq\partial M,\\
0,  &  e\subseteq\partial M.
\end{array}\right.
\end{eqnarray}
We then denote the range of $\nabla$ as  $\mathbf{V}_M$. For $\mathbf{p} \in\mathbf{V}_M$,  as the adjoint operator of $\nabla$ is -$\mathrm{div}$,   we then have the following divergence operator for $M$.
\begin{equation}\label{div_M}
 (\mathrm{div}(\mathbf{p}))_{\tau} = \frac{-1}{A_{\tau}}\sum\limits_{e\prec\tau} \mathbf{p}_{e}\mathrm{sgn}(\tau,e)l_e,
\end{equation}
where $A_{\tau}$ is the area of triangle $\tau$, and $l_e$ is the length of edge $e$.
\subsection{Relaxed Total generalized variation regularization}
According to the above operators \eqref{grad_M} and \eqref{div_M}, we present the following  total generalized variation regularization
$(\mathrm{TGV}^p, p=1,2$ indicating the order information), which contains the first order total variation
regularization $(\mathrm{TGV}^1=\mathrm{TV})$ and the second order relaxed total generalized variation regularization
$(\mathrm{TGV}^2=\mathrm{RTGV})$.

For $\mathbf{u} \in \mathbf{U}_M, \mathbf{v}\in \mathbf{V}_M$,  we then have
\begin{equation}\label{GTV}
\begin{split}
&\mathrm{TGV}^1(\mathbf{u})= \mathrm{TV}(\mathbf{u}) = \|\nabla\mathbf{u}\|_1 = \sum\limits_{e}\sum\limits_{i=1}^{n}|(\nabla\mathbf{u})_{ei}|l_{e},\\
&\mathrm{TGV}^2(\mathbf{u})=\mathrm{RTGV}(\mathbf{u}) = \min_\mathbf{v} \|\nabla\mathbf{u-v}\|_1 + \alpha_0\|\mathrm{div}(\mathbf{v})\|_1\\
        & \ \ \ \ \ \ \ \ \ \ \ \ \ = \min_\mathbf{v}\sum\limits_{e}\sum\limits_{i=1}^{n}|(\nabla\mathbf{u}-\mathbf{v})_{ei}|l_{e}+ \alpha_0\sum\limits_{\tau}\sum\limits_{i=1}^{n}|\mathrm{div}(\mathbf{v})_{\tau i}|A_{\tau}.
\end{split}
\end{equation}

The $\mathrm{TV}$ regularization  has been applied for mesh denoising \cite{Zhang15} and piecewise constant MS mesh segmentation \cite{Zhang12,zhang2018new}. However, for meshes with ambiguous semantic information or irregular  regions (for  the meshes in Fig.~\ref{test_doublebell},\ref{test_pig}), the piecewise constant MS mesh
segmentation methods  may fall into local minima, and have difficulty in getting the better segmentation boundaries. The $\mathrm{RTGV}$ regularization has been applied for mesh denoising \cite{zhang2022}, and is able to recover the first order discontinuity  of a signal and  overcome the stair-case artifacts (false edges) of TV regularization.  In the following,
we focus on $\mathrm{TGV}^p$ regularized  piecewise smooth MS mesh segmentation method to obtain better segmentation boundaries other than the shortest boundaries.

\section{$\mathrm{TGV}^p$ regularized piecewise smooth MS mesh segmentation method}

As mentioned before, for a mesh $M$ with region $\Omega$ to be divided into $\mathbf{K}$ disjoint parts
$\{\Omega_k\}_{k=1}^{\mathbf{K}}$, the piecewise smooth MS model tries to approximate feature function
of $M$ by $\mathbf{g}(x)=\sum\limits_k\mu_k\mathbf{u}+\mathbf{b}$: piecewise constant values $\{\mu_k\}$ and a smooth function $\mathbf{b}$.
\subsection{The feature function $\mathbf{f}$ of  mesh}
A popular way  to define the feature
functions of images and meshes is using the spectral space, which is spanned by the eigenvectors of a Laplacian matrix
\cite{Shi00,Von07,Liu04,Zhang12,zhang2018new}. In the paper, we adopt the high order Laplacian matrix in \cite{zhang2018new}
to construct the feature function $\mathbf{f}$, which involved two-type neighborhood normals $N_{2}$.

Specifically, for a given mesh, the Laplacian matrix $L = [L_{ij}]$ in \cite{zhang2018new} is  given as follows:
\begin{eqnarray}\label{laplacian_matrix}
 L_{ij} =\left\{\begin{array}{ll}
 -w_{ij},  & i\neq j\ \mathrm{and}\  \tau_i\ ,\   \tau_j\   \mathrm{share}\  \mathrm{an}\  \mathrm{edge},\\
 \sum\limits_k w_{i,k},  & i = j \ \mathrm{and} \  k \in N_{1}(\tau_i), \\
 0, & \mathrm{otherwise},
\end{array}\right.
\end{eqnarray}
where
$w_{ij} = l_e\mathrm{exp}(-\frac{d(\tau_i,\tau_j)}{\overline{d}}),$
 $\overline{d}$ being the average of $d(\tau_i,\tau_j)$ over all edges, and $d(\tau_i,\tau_j)$ denotes the error between normals.
 In Fig.~\ref{test_doublebell},\ref{test_pig}, we present the feature functions of the two meshes.

\subsection{Piecewise smooth MS mesh segmentation method }
By introducing a multi-continuous label $\mathbf{u}\in C_{\mathbf{u}} = \{\mathbf{u_{\tau}}| u_{\tau,k} \geq 0, \sum\limits_{k = 1}^{\mathbf{K}} u_{\tau,k} = 1, \ \ \forall\tau \}\subset\mathbf{U}_M$,
we then obtain the following piecewise smooth MS mesh segmentation optimization problem.
\begin{equation}\label{ps_MS_Mesh}
\begin{split}
\min_{\mathbf{u}\in C_{\mathbf{u}},\mathbf{b},\mu}& E(\mathbf{u},\mathbf{b},\mu)\\
& = \mathrm{TGV}^p(\mathbf{u}) + \frac{\beta}{2}\|\triangle\mathbf{b}\|^2_{\mathbf{U}_M}+\frac{\eta}{2}\|\mathbf{b}\|^2_{\mathbf{U}_M} + \frac{\alpha}{2}\langle\mathbf{u},s(\mathbf{f},\mathbf{b},\mu)\rangle_{\mathbf{U}_M},
\end{split}
\end{equation}
where the second term is to get the smooth parts of $\mathbf{f}$, the third term is used to stabilize the  $\mathbf{b}$  sub problem, the last term is the similarity term, and $s(\mathbf{f},\mathbf{b},\mu)=\{s_{\tau}(\mathbf{f},\mathbf{b},\mu)\}$ with $s_{\tau}(\mathbf{f},\mathbf{b},\mu)=\{\|\mathbf{f}_\tau-\mathbf{b}_\tau-\mu_k\|^2\}_{k=1}^{\mathbf{K}}$,
$\mathbf{f}$ is the  eigenvectors of the smallest $\mathbf{K}-1$ eigenvalues of the Laplace matrix, and $\beta, \eta,\alpha$ are positive parameters.
As \eqref{ps_MS_Mesh} is non-convex, the solution is not unique. Motivated by the techniques of \cite{jung2017piecewise,li2020tv}, it is still available to show the existence of minimizers of \eqref{ps_MS_Mesh}.
\begin{theorem}
Let $\Omega$ be a bounded subset of $R^3$  with a Lipschitz boundary. Assume that $\mathbf{f}$ and $\{\mu_k\}_{k=1}^{\mathbf{K}}$ are bounded. Then, for $\beta,\eta,\alpha > 0,$ there exists a minimizer $(\mathbf{u}^\ast,\mathbf{b}^\ast,\mu^\ast)$ for problem \eqref{ps_MS_Mesh}.  (The proof is shown in the APPENDIX)
\end{theorem}

\subsection{Alternating minimization method (AMM) for solving \eqref{ps_MS_Mesh}}

 Given  the following function
\begin{eqnarray}\chi (\mathbf{u})=\left\{\begin{array}{ll}
0,  & \mathbf{u} \in C_{\mathbf{u}}\\
+\infty,  & \mathbf{u} \notin C_{\mathbf{u}},
\end{array}\right.,
\end{eqnarray}
the minimization problem \eqref{ps_MS_Mesh} can be solved alternately by  the following  two optimization problems.
\begin{itemize}
\item For fixed $\mathbf{\mu}$: $\mathbf{u},\mathbf{b}$ can be obtained by solving

\begin{equation}\label{seg_ub}
\begin{split}
\min_{\mathbf{u},\mathbf{b},\mu} &\mathrm{TGV}^p(\mathbf{u}) + \frac{\beta}{2}\|\triangle\mathbf{b}\|^2_{\mathbf{U}_M}+\frac{\eta}{2}\|\mathbf{b}\|^2_{\mathbf{U}_M}
+ \frac{\alpha}{2}\langle\mathbf{u},s(\mathbf{f},\mathbf{b},\mu)\rangle_{\mathbf{U}_M} + \chi (\mathbf{u}),
\end{split}
\end{equation}

which is non-differentiable, and can be efficiently solved by augmented Lagrangian method (ALM) or alternating direction
multiplier method (ADMM) \cite{Wu10SIAM,Yang10,Michael10,Boyd11}; see section~\ref{solve_ub} for details.\\

\item For fixed $\mathbf{u},\mathbf{b}$: $\mathbf{\mu}$ can be obtained by solving
\begin{equation*}
\begin{split}
\min\limits_{\mu}
 \frac{\alpha}{2}\langle\mathbf{u},s(\mathbf{f},\mathbf{b},\mu)\rangle_{\mathbf{U}_M},
\end{split}
\end{equation*}
and is exactly
\begin{equation}\label{seg_mu}
\mathbf{\mu}_{k} = \frac{\sum\limits_{\tau}u_{k,\tau}(\mathbf{f}_{\tau}-\mathbf{b}_{\tau})A_{\tau}}{\sum\limits_{\tau}u_{k,\tau}A_{\tau}},\ \ \ \  k = 0, 1, \cdots, \mathbf{K}.
\end{equation}
\end{itemize}

In our implementation, the alternating minimization method is listed in Algorithm~\ref{algorithm1}.
\begin{algorithm}{htbp}
\caption{\label{algorithm1} AMM for solving  \eqref{ps_MS_Mesh} }
\begin{itemize}
\item[1.] \textbf{Initialization}:
\begin{itemize}
\item[$\bullet$] \textbf{$\mu^{-1}:$} computed according to the technique in \cite{zhang2018new};
\item[$\bullet$] $\mathbf{u}^{-1}:$ obtained by \textbf{$\mu^0$};
\item[$\bullet$]
    $\mathbf{v}^{-1} = 0,\  \mathbf{b}^{-1} = 0$, \  $\mathbf{p}^{-1} = 0, \ \mathbf{q}^{-1} = 0, \ \mathbf{z}^{-1} = 0$,\\
    $\lambda_{\mathbf{p}}^{-1} = 0,\  \lambda_{\mathbf{q}}^{-1} = 0,\  \lambda_{\mathbf{z}}^{-1} = 0,\  l = -1$;
\end{itemize}
\item[2.] \textbf{Repeat}
\begin{itemize}
\item[$\bullet$] For fixed $\mathbf{\mu}^{l-1}$,  computing \\  $(\mathbf{u}^{l},\mathbf{v}^{l}, \mathbf{b}^{l},\mathbf{p}^{l},\mathbf{ q}^{l},\mathbf{z}^{l},\lambda_{\mathbf{p}}^{l},\lambda_{\mathbf{q}}^{l},\lambda_{\mathbf{z}}^{l} )$ by solving \eqref{seg_ub} through  Algorithm~\ref{algorithm2};
\item[$\bullet$] For fixed ($\mathbf{u}^{l},\mathbf{b}^{l}$), computing $\mathbf{\mu}^{l}$ from \eqref{seg_mu};
\end{itemize}
\item[]\textbf{Until}\ ($\|\mathbf{u}^{n+1}-\mathbf{u}^{n}\|^2_{\mathbf{U}_{M}}<10^{-5}$).
\item[3.] \textbf{Classify $\mathbf{u}$} by the method in \cite{Lellmann11}.
\end{itemize}
\end{algorithm}

\section{Algorithm details  for solving  (15) } \label{solve_ub}
Generally, in the following, we only present the details for solving \eqref{seg_ub} with $p = 2$ (RTGV regularization).
By introducing three auxiliary variables $\mathbf{p}\in \mathbf{V}_{M}$ and $\mathbf{q}, \ \mathbf{z} \in \mathbf{U}_{M}$. The problem \eqref{seg_ub} with $p=2$ can be further written as
\begin{equation}\label{seg_ubz}
\begin{split}
\min_{\substack{\mathbf{u},\mathbf{b}\in\mathbf{U}_M,\mathbf{v}\in\mathbf{V}_M,\\ \mathbf{p}\in\mathbf{V}_M,\mathbf{q}, \mathbf{z}\in\mathbf{U}_M}}
  &\|\mathbf{p}\|_{1}+\alpha _{0}\|\mathbf{q}\|_{1}+
  \frac{\beta}{2}\|\mathbf{\triangle b}\|^{2}_{\mathbf{U}_M}+\frac{\eta}{2}\|\mathbf{b}\|^{2}_{\mathbf{U}_M}
 +\frac{\alpha}{2}(\mathbf{z},\mathbf{s}(\mathbf{f},\mathbf{b},\mathbf{\mu}))_{\mathbf{U}_M} +\chi (\mathbf{z}),\\
 &\mathrm{s.t.} \ \ \ \mathbf{p}=\mathbf{\nabla u}-\mathbf{v},\ \mathbf{q}=\mathrm{div}(\mathbf{v}),\ \ \mathbf{z}=\mathbf{u}.
\end{split}
\end{equation}

 To solve \eqref{seg_ubz} effectively, we define
the following augmented Lagrangian functional:
\begin{equation}\label{seg_alm}
\begin{split}
\mathcal{L}&(\mathbf{u,v,b,p,q,z; \lambda_{p},\lambda_{q}, \lambda_{z}})
 = \|\mathbf{p}\|_{1}+\alpha _{0}\|\mathbf{q}\|_{1}+\frac{\beta}{2}\|\mathbf{\triangle b}\|^{2}_{\mathbf{U}_M}+\frac{\eta}{2}\|\mathbf{b}\|^{2}_{\mathbf{U}_M} \\ &+\frac{\alpha}{2}(\mathbf{z},\mathbf{s}(\mathbf{f},\mathbf{b},\mathbf{\mu}))_{\mathbf{U}_M}+ \chi (\mathbf{z})
+(\lambda_{\mathbf{p}},\mathbf{p-(\nabla u-v)})_{\mathbf{V}_{M}}\\
&+\frac{{r_{\textbf{p}}}}{2}\|\mathbf{p-(\nabla u-v)}\|^{2}_{\mathbf{V}_M}
+(\lambda_{\mathbf{q}},\mathbf{q}-\mathrm{div}(\mathbf{v}))_{\mathbf{U}_{M}}
+\frac{{r_{\textbf{q}}}}{2}\|\mathbf{q}-\mathrm{div}(\mathbf{v})\|^{2}_{\mathbf{U}_M}\\
&+(\lambda_{\mathbf{z}},\mathbf{z-u})_{\mathbf{U}_{M}}+\frac{{r_{\textbf{z}}}}{2}\|\mathbf{z-u}\|^{2}_{\mathbf{U}_M},
\end{split}
\end{equation}
where $r_{\textbf{p}},r_{\textbf{q}} $ and $ r_{\textbf{z}} $ are positive parameters.

The solution of \eqref{seg_alm} is equivalent to the following saddle-point problem
\begin{equation}\label{min_max}
\max_{\lambda_\mathbf{p}, \lambda_\mathbf{q},\lambda_\mathbf{z}}\min_{\substack{\mathbf{u},\mathbf{v, b,}\\ \mathbf{p, q,z}}}\mathcal{L}(\mathbf{u,v,b,p,q,z};\lambda_\mathbf{p}, \lambda_\mathbf{q},\lambda_\mathbf{z}),
\end{equation}
which can be iteratively solved by   splitting \eqref{min_max} into the following several subproblems.
\subsection{Sub-minimizations with respect to $\mathbf{u}, \mathbf{v}, \mathbf{b} $}

The $\mathbf{u, v, b}$ sub-problems are reformulated as follows
\begin{equation}\label{seg_subu}
\begin{split}
 \min\limits_{\mathbf{u}\in\mathbf{U}_{M}} (\lambda_{\mathbf{p}},\mathbf{-\nabla\mathbf{u}})_{\mathbf{V}_M}+(\lambda_{\mathbf{z}},-\mathbf{u})_{\mathbf{U}_M}& +\frac{{r_{\textbf{p}}}}{2}\|\mathbf{p-(\nabla \mathbf{u}}- \mathbf{v})\|^{2}_{\mathbf{V}_M}
 + \frac{{r_{\textbf{z}}}}{2}\|\mathbf{z-u}\|^{2}_{\mathbf{U}_M},
\end{split}
\end{equation}
\begin{equation}\label{seg_subv}
\begin{split}
 \min\limits_{\mathbf{v}\in\mathbf{Q}_{M}} (\lambda_{\mathbf{q}},-\mathrm{div}(\mathbf{v}))_{\mathbf{U}_M}&+\frac{{r_{\textbf{q}}}}{2}\|\mathbf{q}-\mathrm{div}(\mathbf{v})\|^{2}_{\mathbf{U}_M}
+(\lambda_{\mathbf{p}},\mathbf{v})_{\mathbf{V}_{M}}
+\frac{{r_{\textbf{p}}}}{2}\|\mathbf{p-(\nabla u-v)}\|^{2}_{\mathbf{V}_M},
\end{split}
\end{equation}
\begin{equation}\label{seg_subb}
 \min\limits_{\mathbf{b}\in\mathbf{U}_{M}}\frac{\alpha}{2}(\mathbf{z}, \mathbf{s}(\mathbf{f},\mathbf{b},\mathbf{\mu}))_{\mathbf{U}_{M}}+\frac{\beta}{2}\|\mathbf{\triangle b}\|^{2}_{\mathbf{U}_{M}}+\frac{\eta}{2}\|\mathbf{b}\|^{2}_{\mathbf{U}_{M}}.\  \  \ \ \ \ \ \ \ \ \ \ \ \ \ \ \ \ \ \ \ \
\end{equation}
The above sub-problems are quadratic programming problems, and can be  solved by various numerical packages, such as MKL, Taucs and Eigen.

\subsection{Sub-minimizations with respect to $\mathbf{p}, \mathbf{q}, \mathbf{z} $}
Specifically, the $\mathbf{p, q, z}$ sub-problems  can be reformulated as follows and  solved with the closed form solutions.
\begin{itemize}
\item For $\mathbf{p}$ sub-problem, we have
\begin{equation}\label{seg_subp}
\min\limits_{\mathbf{p}\in\mathbf{V}_{M}}\|\mathbf{p}\|_{1}+(\lambda_{\mathbf{p}},\mathbf{p})_{\mathbf{V}_{M}}+
\frac{{r_{\textbf{p}}}}{2}\|\mathbf{p-(\nabla u-v)}\|^{2}_{\mathbf{V}_M}.
\end{equation}
It has the following closed form solution
\begin{eqnarray}\label{seg_subp_solution}
\forall e, \mathbf{p}_{e}=\left\{\begin{array}{ll}
(1-\frac{1}{{r_{\mathbf{p}}}|\mathbf{w}_{e}|})\mathbf{w}_e,  & |\mathbf{w}_{e}|>\frac{1}{{r_{\mathbf{p}}}},\\
0,  & |\mathbf{w}_{e}|\leq\frac{1}{{r_{\mathbf{p}}}},
\end{array}\right.
\end{eqnarray}
where $\mathbf{w}_e = (\nabla \mathbf{u} - \mathbf{v} -\frac{\lambda_{\mathbf{p}}}{r_{\mathbf{p}}})|_e$.
\item For $\mathbf{q}$ sub-problem, we have
\begin{equation}\label{seg_subq}
\min\limits_{\mathbf{q}\in\mathbf{U}_{M}} \alpha _{0}\|\mathbf{q}\|_{1}+(\lambda_{\mathbf{q}},\mathbf{q})_{\mathbf{U}_{M}}+
\frac{{r_{\textbf{q}}}}{2}\|\mathbf{q}-\mathrm{div}(\mathbf{v})\|^{2}_{\mathbf{U}_M},
\end{equation}
which has the following closed form solution
\begin{eqnarray}\label{seg_subq_solution}
\forall \tau, \mathbf{q}_{\tau}=\left\{\begin{array}{ll}
(1-\frac{\alpha _{0}}{{r_{\mathbf{q}}}|\mathbf{c}_{\tau}|})\mathbf{c}_\tau,  & |\mathbf{c}_{\tau}|>\frac{\alpha_0}{{r_{\mathbf{q}}}},\\
0,  & |\mathbf{c}_{\tau}|\leq\frac{\alpha_0}{{r_{\mathbf{q}}}},
\end{array}\right.
\end{eqnarray}
where $\mathbf{c}_\tau = \mathrm{div}(\mathbf{v})-\frac{\lambda_{\mathbf{q}}}{r_{\mathbf{q}}}$.
\item For $\mathbf{z}$ sub-problem, we get
\begin{equation}\label{seg_subz}
\min\limits_{\mathbf{z}\in\mathbf{U}_{M}}\alpha(\mathbf{z}, \mathbf{s}(\mathbf{f},\mathbf{\mu},\mathbf{b}))_{\mathbf{U}_M} + \chi\mathbf{(z)}+(\lambda_{\mathbf{z}},\mathbf{z})_{\mathbf{U}_M}+\frac{{r_{\mathbf{z}}}}{2}\|\mathbf{z-u}\|^{2}_{\mathbf{U}_M}.
\end{equation}
It can be solved through
\begin{equation}\label{seg_subz_solution}
\mathbf{z} = \mathrm{Proj}_{C_{\mathbf{u}}}(\mathbf{u}-\frac{\alpha \mathbf{s}(\mathbf{f},\mathbf{b},\mathbf{\mu})) +\lambda_{\mathbf{z}}}{r_{\mathbf{z}}}),
\end{equation}
which can be calculated via Michelot's algorithm \cite{Michelot86}.
\end{itemize}

The whole  algorithm for solving \eqref{min_max} is listed in Algorithm~\ref{algorithm2}.
\begin{algorithm}[htbp]
\caption{\label{algorithm2} ADMM for solving  \eqref{min_max} }
\begin{itemize}
\item[1.] \textbf{Initialization}:
\begin{itemize}
\item[$\bullet$] $\mathbf{u}^{l,0} = \mathbf{u}^{l-1}, \mathbf{v}^{l,0} = \mathbf{v}^{l-1},\mathbf{b}^{l,0} = \mathbf{b}^{l-1},
\mathbf{p}^{l,0} = \mathbf{p}^{l-1},\mathbf{q}^{l,0} = \mathbf{q}^{l-1}, \mathbf{z}^{l,0} = \mathbf{z}^{l-1};$
\item[$\bullet$] $\mathbf{\lambda}^{l,0}_\mathbf{p} = \mathbf{\lambda}^{l}_\mathbf{p}, \mathbf{\lambda}^{l,0}_\mathbf{q} = \mathbf{\lambda}^{l}_\mathbf{q},  \mathbf{\lambda}^{l,0}_\mathbf{z} = \mathbf{\lambda}^{l}_\mathbf{z}$, $K = 5$;
\end{itemize}
\item[2.] For $k = 1,2,\cdots,K$
\begin{itemize}
\item[$\bullet$] $\mathbf{z}$-subproblem: For fixed $(\mathbf{\lambda}^{l,k}_\mathbf{z}, \mathbf{u}^{l,k}, \mathbf{b}^{l,k})$,  compute $\mathbf{z}^{l,k+1}$ from \eqref{seg_subz_solution};
\item[$\bullet$] $\mathbf{u}$-subproblem: For fixed $(\mathbf{\lambda}^{l,k}_\mathbf{p}, \mathbf{\lambda}^{l,k}_\mathbf{z},  \mathbf{p}^{l,k} , \mathbf{v}^{l,k}, \mathbf{z}^{k,k+1})$, compute $\mathbf{u}^{l,k+1}$ by solving \eqref{seg_subu};
\item[$\bullet$] $\mathbf{v}$-subproblem: For fixed $(\mathbf{\lambda}^{l,k}_\mathbf{p}, \mathbf{\lambda}^{l,k}_\mathbf{q},  \mathbf{p}^{l,k} , \mathbf{q}^{l,k}, \mathbf{u}^{l,k+1})$, compute $\mathbf{v}^{l,k+1}$ by solving \eqref{seg_subv};
\item[$\bullet$] $\mathbf{b}$-subproblem: For fixed $(\mathbf{z}^{l,k+1}),$  compute $\mathbf{b}^{l,k+1}$ by solving \eqref{seg_subb};
\item[$\bullet$] $\mathbf{p}$-subproblem: For fixed $(\mathbf{\lambda}^{l+1,k}_\mathbf{p}, \mathbf{v}^{l,k+1}, \mathbf{u}^{l,k+1}),$  compute $\mathbf{p}^{l,k+1}$ from \eqref{seg_subp_solution};
\item[$\bullet$] $\mathbf{q}$-subproblem: For fixed $(\mathbf{\lambda}^{l+1,k}_\mathbf{q}, \mathbf{v}^{l,k+1}),$  compute $\mathbf{q}^{l,k+1}$ from \eqref{seg_subq_solution};
\item[$\bullet$] Update Lagrange multipliers;
\item[]$\lambda^{l,k+1}_\mathbf{p} = \lambda^{l,k}_\mathbf{p} + r_\mathbf{p}(\mathbf{p}^{l,k+1}+\mathbf{v}^{l,k+1}-\nabla \mathbf{u}^{l,k+1})$;
\item[]$\lambda^{l,k+1}_\mathbf{q} = \lambda^{l,k}_\mathbf{q} + r_\mathbf{q}(\mathbf{q}^{l,k+1}-\mathrm{div}(\mathbf{v}^{l,k+1}))$;
\item[]$\lambda^{l,k+1}_\mathbf{z} = \lambda^{l,k}_\mathbf{z} + r_\mathbf{z}(\mathbf{u}^{l,k+1}- \mathbf{z}^{l,k+1})$;
\end{itemize}
\item[3.]$\mathbf{u}^{l} = \mathbf{u}^{l,K}, \mathbf{v}^{l} = \mathbf{v}^{l,K}, \mathbf{b}^{l} = \mathbf{b}^{l,K},$
$\mathbf{p}^{l} = \mathbf{p}^{l,K}, \mathbf{q}^{l} = \mathbf{q}^{l,K}, \mathbf{z}^{l} = \mathbf{z}^{l,K},$
\item[]$\mathbf{\lambda}^{l}_\mathbf{p} = \mathbf{\lambda}^{l,K}_\mathbf{p}, \mathbf{\lambda}^{l}_\mathbf{q} = \mathbf{\lambda}^{l,K}_\mathbf{q},  \mathbf{\lambda}^{l}_\mathbf{z} = \mathbf{\lambda}^{l,K}_\mathbf{z}.$
\end{itemize}
\end{algorithm}
As the MS segmentation optimization problem \eqref{seg_ubz} is non-convex, it is difficult to show the convergence of our algorithm. Similar to  the theorems given in \cite{jung2017piecewise,li2020tv}, we can only get the following partial convergence theorem.
\begin{theorem}
  Supposed $\mathbf{u}^{l}-\mathbf{u}^{l-1}\rightarrow 0, \mathbf{v}^{l}-\mathbf{v}^{l-1}\rightarrow 0, \lambda_\mathbf{p}^{l}-\lambda_\mathbf{p}^{l-1}\rightarrow 0, \lambda_\mathbf{q}^{l}-\lambda_\mathbf{q}^{l-1} \rightarrow 0,  \lambda_\mathbf{z}^{l}-\lambda_\mathbf{z}^{l-1}\rightarrow 0, $ in Algorithm 1. If the sequence $\{(\mathbf{u}^l,\mathbf{v}^l,\mathbf{b}^l,\mu^l,\mathbf{p}^l,\mathbf{q}^l, $ $\mathbf{z}^l,\lambda_\mathbf{p}^l,\lambda_\mathbf{q}^l,\lambda_\mathbf{z}^l)\}$   generated by the Algorithm 1 converges to a limit point $(\mathbf{u}^\ast,\mathbf{v}^\ast,\mathbf{b}^\ast,$ $\mu^\ast,\mathbf{p}^\ast,\mathbf{q}^\ast,\mathbf{z}^\ast,\lambda_\mathbf{p}^\ast,\lambda_\mathbf{q}^\ast,\lambda_\mathbf{z}^\ast)$, then the limit point satisfies the following Karush-Kuhn-Tucker $(\mathrm{KKT})$ conditions of \eqref{seg_ubz}:
  \begin{eqnarray}
   \left\{\begin{array}{l}
     -\lambda_\mathbf{z}^\ast+ \mathrm{div}(\lambda_\mathbf{p}^\ast)=0,\\
       \lambda_\mathbf{p}^\ast +\nabla(\lambda_\mathbf{q}^\ast)  = 0,\\
  (\beta \Delta^2 +\eta\mathcal{I}+\alpha\mathcal{I})\mathbf{b}^\ast + \alpha<-(\mathbf{f}-\mu^\ast),\mathbf{z}^\ast> = 0,\\
   \alpha<\mu^\ast-(\mathbf{f}-\mathbf{b}^\ast),\mathbf{u}^\ast> = 0,\\
   \partial(\|\mathbf{p}^\ast\|_1)+ \lambda_\mathbf{p}^\ast = 0,\\ \alpha_0\partial(\|\mathbf{q}^\ast\|_1)+ \lambda_\mathbf{q}^\ast = 0,\\
 \alpha s(\mathbf{f},\mathbf{b}^\ast,\mu^\ast) + \lambda_\mathbf{z}^\ast +\partial\chi(\mathbf{z}^\ast) = 0,\\
 \mathbf{p}^\ast-(\nabla \mathbf{u}^\ast-\mathbf{v}^\ast)=0,\\ \mathbf{q}^\ast-\mathrm{div}(\mathbf{v}^\ast)=0, \\\mathbf{z}^\ast-\mathbf{u}^\ast=0,
  \end{array}\right.
\end{eqnarray}
(The proof is shown in the APPENDIX)
\end{theorem}

\section{\label{resultdiscussion}Experimental results and discussions}
In this section, we present  experiments of our piecewise smooth MS segmentation method on a wide variety of mesh
surfaces, including of the Princeton Segmentation Benchmark \cite{Chen09} and some other surfaces (meshes in Fig.~\ref{complex_result}). Our segmentation methods
are denoted as ``$\mathrm{PSMS}$" (piecewise smooth MS method based on the TV regularization and ``$\mathrm{GPSMS}$" (piecewise smooth MS method based on the RTGV regularization),
respectively. The experiments of our algorithms and the method in \cite{zhang2018new} are conducted  using Microsoft
Visual Studio 2010 on a  desktop with Intel(R) Core(TM) i7-8850 CPU @2.60GHz and 16GB memory and all mesh surfaces
were rendered using flat shading.

We discuss our algorithm from several aspects, such as choices of parameters  and comparisons
to existing techniques by visual effects and quantitative errors. Specifically, we compare our segmentation method with
Isoline Cut \cite{Oscar12},  WCSeg \cite{Kaick14}, PCMS (piecewise constant MS) \cite{zhang2018new}, Spectral $L_0$
\cite{tong2018spectral}, and two learning based methods SB19 \cite{Kalogerakis10}.
  All the segmentation results are provided by  the authors. In addition, for evaluating segmentation results, we
  use the four evaluation metrics in \cite{Chen09}.

\subsection{Parameters and the number of segments}
There are seven parameters in our algorithm: $\alpha, \beta, \alpha_0, $ $\eta, r_\mathbf{p}, r_\mathbf{q}$ and
$r_{\mathbf{z}}$. Therein $\alpha_0, \alpha, \beta, \eta$ are  variational model parameters;
$r_\mathbf{p},r_\mathbf{q},r_\mathbf{z}$ are optimization algorithmic parameters. The parameters
$\eta, r_\mathbf{p}, r_\mathbf{q}$ and $r_{\mathbf{z}}$ can be  fixed by
$\eta = 0.00001, r_\mathbf{p} = 1,  r_\mathbf{q} = 1, r_{\mathbf{z}} = 100 $. Moreover,  $\alpha$, $\beta$ and
$\alpha_0$ all have an impact on  the segmentation results. According to lots of experimental tests, the parameter settings of
$\alpha$, $\beta$ and $\alpha_0$ are summarized  as follows.

\begin{figure}[htbp]
\centering\includegraphics*[width=4.5in]{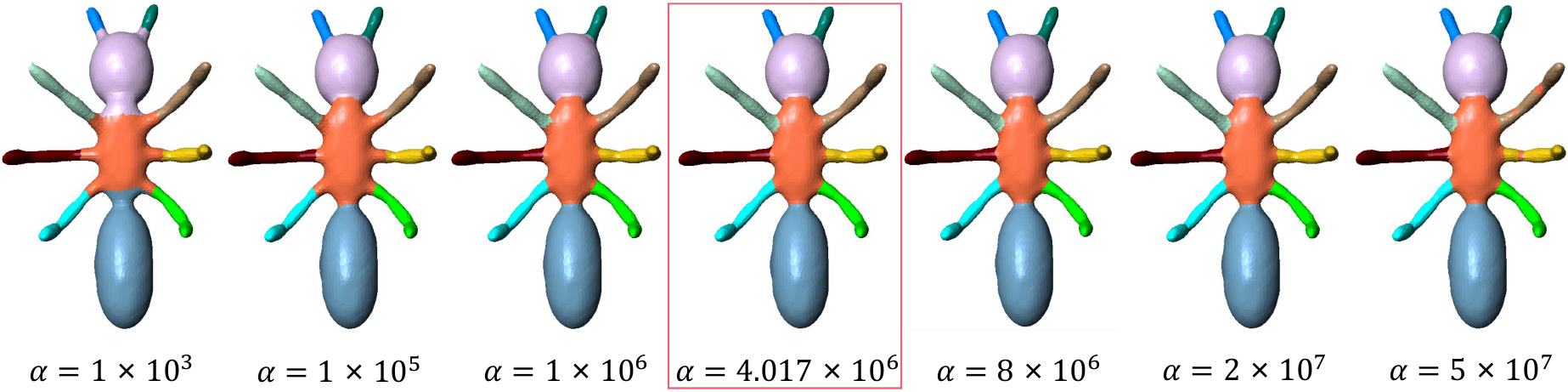}
\caption{{
The segmentation results of different $\alpha$ by our $\mathrm{ GPSMS}$ method  with other parameters fixed.
}
}\label{para_alpha}
\end{figure}

Firstly, the parameter $\alpha$ affects the segmentation boundary.  Fig.~\ref{para_alpha} presents the impact of the parameter $\alpha$ on the results. As shown, the larger the $\alpha$ is, the better the segmentation
boundary is. However, when the
$\alpha$ is too large, the algorithm may over segment the mesh ( see the result with $\alpha = 5\times 10^7$
in Fig.~\ref{para_alpha}). In addition,  according to our tests,  the number of segment $\mathbf{K}$,  the similarity term and the
regularization term in \eqref{ps_MS_Mesh} all affect the selection of  $\alpha$.
 based on these facts, we then provide the following  empirical formula to estimate the $\alpha$.
\begin{equation}\label{alpha_formula}
\alpha = 2\times\mathbf{K}\times \frac{(\mathrm{TGV})(\mathbf{u^{-1}})}{(\mathbf{u^{-1}}, \mathbf{s}(\mathbf{f, b^{-1}},\mathbf{\mu}^{-1}))_{\mathbf{U}_M}}.
\end{equation}
Experiments demonstrate that the  empirical formula works well for many situations. In Fig.~\ref{para_alpha},  we
present the result produced with $\alpha$ computed by \eqref{alpha_formula} (see the results marked by the red frame).
 As seen, the segmentation result with $\alpha = 4.017\times 10^6$  is quite satisfying and the segmentation boundaries
 are consistent with human perception.

\begin{figure}[htbp]
\centering\includegraphics*[width=4.5in]{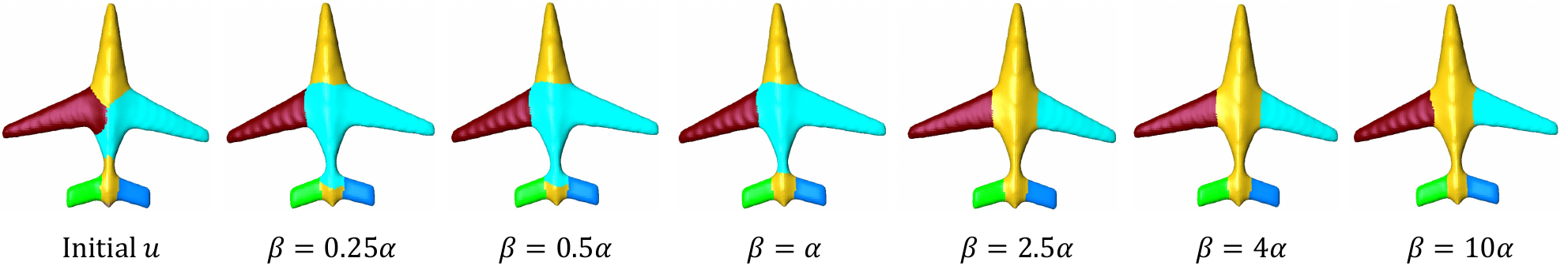}
\centering\includegraphics*[width=4.5in]{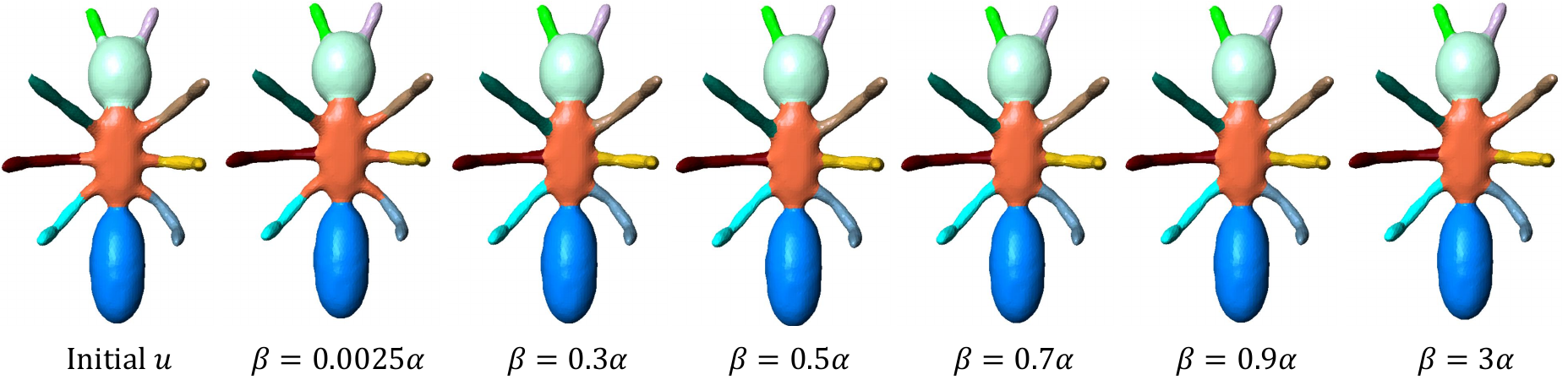}
\caption{{
The segmentation results  of different $\beta$ by our $\mathrm{ GPSMS}$ method  with other parameters fixed.
}
}\label{para_beta}
\end{figure}

Secondly, the parameter $\beta$ plays an important impact on the results. Experiment indicates that
  $\beta$ is associated with  $\alpha$. According to the tests,  there exist a range of $\beta$ for our
 algorithm giving good segmentation results. Specifically, if the initial segmentation boundary is redundant (see the first row in Fig.~\ref{para_beta}), $\beta$ should be larger than $\alpha$.
If the initial segmentation boundary is far from the optimal boundary (see the second row in Fig.~\ref{para_beta}),
$\beta$ should be smaller than $\alpha$. We then have $\beta\in[0.2\alpha, 10\alpha]$. Fig.~\ref{para_beta} presents
the results of different $\beta$ with other parameters fixed, which is consistent with the estimated interval.

\begin{figure}[htbp]
\centering\includegraphics*[width=4.5in]{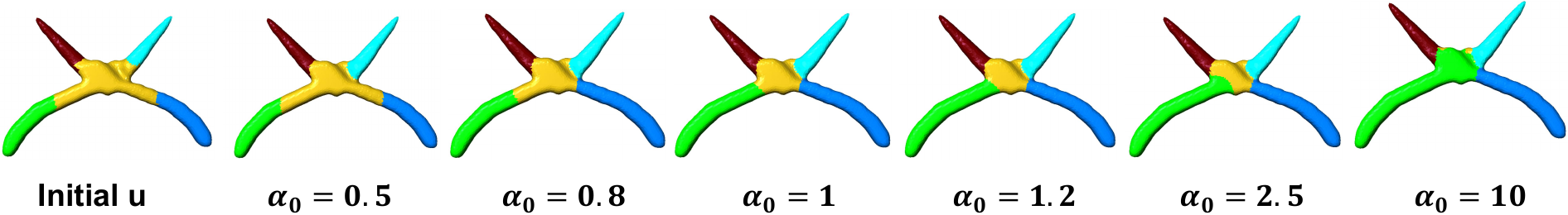}
\caption{
The segmentation results of different $\alpha_0$ by our $\mathrm{ GPSMS}$ method  with other parameters fixed.
}\label{test_alpha0}
\end{figure}

Finally, $\alpha_0$ also affect on the segmentation result, especially for meshes with poor quality. Fig.~\ref{test_alpha0} show the results of different $\alpha_0$
 with other parameters. As seen, whether $\alpha_0$ is too small or too large, the segmentation boundaries are not satisfying. Because of the variety of mesh surfaces and the non-convexity
of MS model, it is difficult to compute the parameter by a formula. However, according to the tests,  $\alpha_0$ can be set by  the range $[1, 6]$.

\begin{remark}
The number of segment $\mathbf{K}$ is affected by the geometric structure and semantic information of the mesh. It is difficult to find an empirical formula suitable for all mesh surfaces. In the paper, $\mathbf{K}$ is set  manually.
\end{remark}
\subsection{Comparison to the piecewise constant MS method (PCMS) \cite{zhang2018new} }

Since both our methods (including PSMS and GPSMS) and piecewise constant MS method (PCMS) \cite{zhang2018new}
are based on  the Mumford-Shah segmentation model,
we would like to compare the three methods in more details, including the two-region segmentation results and the multi-region segmentation results.

\subsubsection{Two-region segmentation}
\begin{figure}[htbp]
\centering\includegraphics*[width=4.5in]{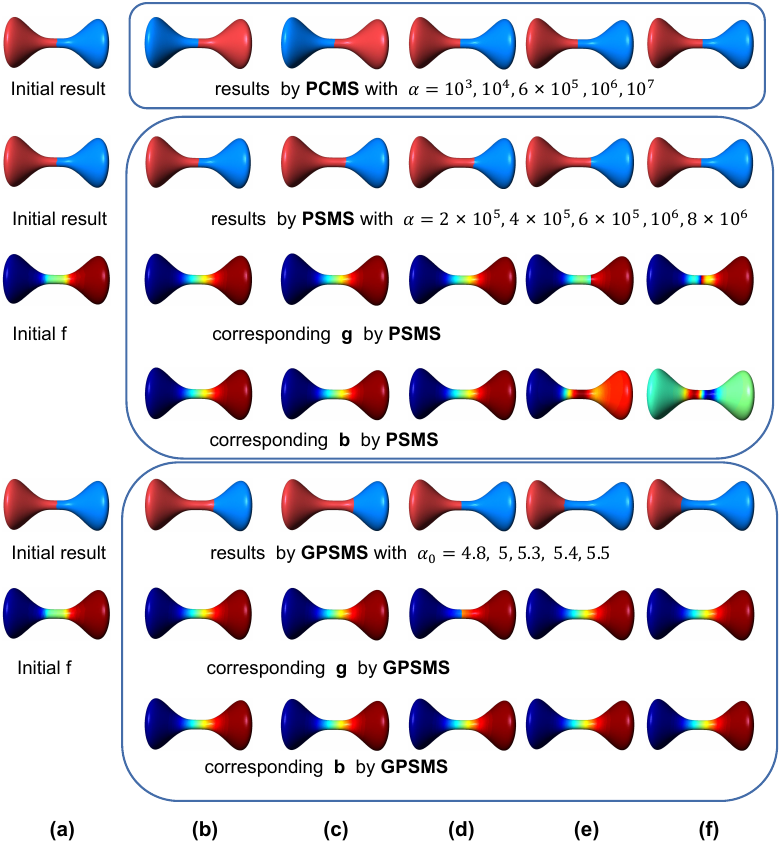}
\caption{
Comparisons of two-region segmentation results for Doublebell mesh by the $\mathrm{PCMS}$ method (the first row), the $\mathrm{PSMS}$ method (the second row), and the $\mathrm{GPSMS}$ method (the fifth row).
}\label{test_doublebell}
\end{figure}

\begin{figure}[htbp]
\centering\includegraphics*[width=4.5in]{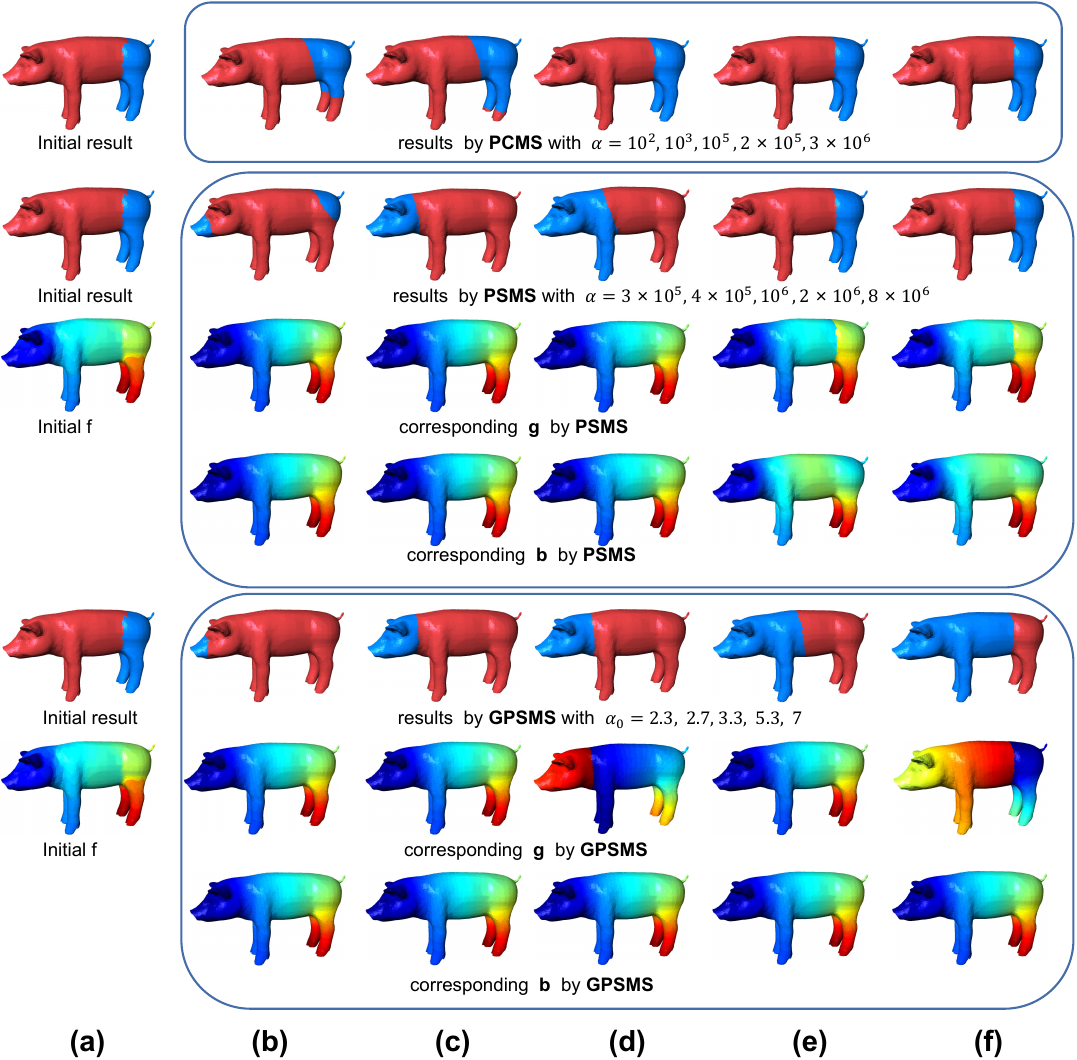}
\caption{
Comparisons of two-region segmentation results for Pig mesh by the $\mathrm{PCMS}$ method (the first row), the $\mathrm{PSMS}$ method (the second row), and the $\mathrm{GPSMS}$ method (the fifth row).
}\label{test_pig}
\end{figure}

We first present the two-region segmentation results by the three methods and the corresponding $\mathbf{g, b}$ by our methods in Fig.~\ref{test_doublebell} and \ref{test_pig}.

As can be seen, for the Doublebell mesh in Fig.~\ref{test_doublebell}, the PCMS method produces results around  the initialization no matter how to adjust the parameter, which is just the discontinuous position of the feature function (see the results in the first row of Fig.~\ref{test_doublebell}).
By introducing the smooth function $\mathbf{b}$, the PSMS method can capture more structures of the mesh, and get results better than that of the PCMS method to a certain extent. However, the segmentation boundaries of the PSMS method are unitary and not optimal (see the results in the second row of Fig.~\ref{test_doublebell}, especially for the result in column (e)). By comparison, our GPSMS method exhibits the ability to acquire the different features of a mesh by using the RTGV regularization, which can identify the high order discontinuity of the feature function. Therefore, with appropriate parameter $\alpha_0$, the GPSMS method can  get a  variety of  solutions  consistent with the human cognition (see the results in the fifth row of Fig.~\ref{test_doublebell}).

In Fig.~\ref{test_pig}, we further test the three methods on the Pig mesh for two-region segmentation. Similarly, the PCMS method can capture the discontinuity of the feature function effectively, and thus are affected severely by the initialization. The PSMS method can improve the  results to some extent (see the results in the second row of Fig.~\ref{test_pig}, especially for the result in column (c)). However, the role of smooth function $\mathbf{b}$ is limited. By comparison, our GPSMS method not only get the discontinuity , but also  characterize the high order discontinuous structures of the feature function effectively. The corresponding  segmentation results are with better boundaries (see the results in the fifth row of Fig.~\ref{test_pig}, especially for the results in columns (b),(c),(d)).
\subsubsection{Multi-region segmentation}
We then present the multi-region segmentation results by the three methods.

\begin{figure}[htbp]
\centering\includegraphics*[width=4.5in]{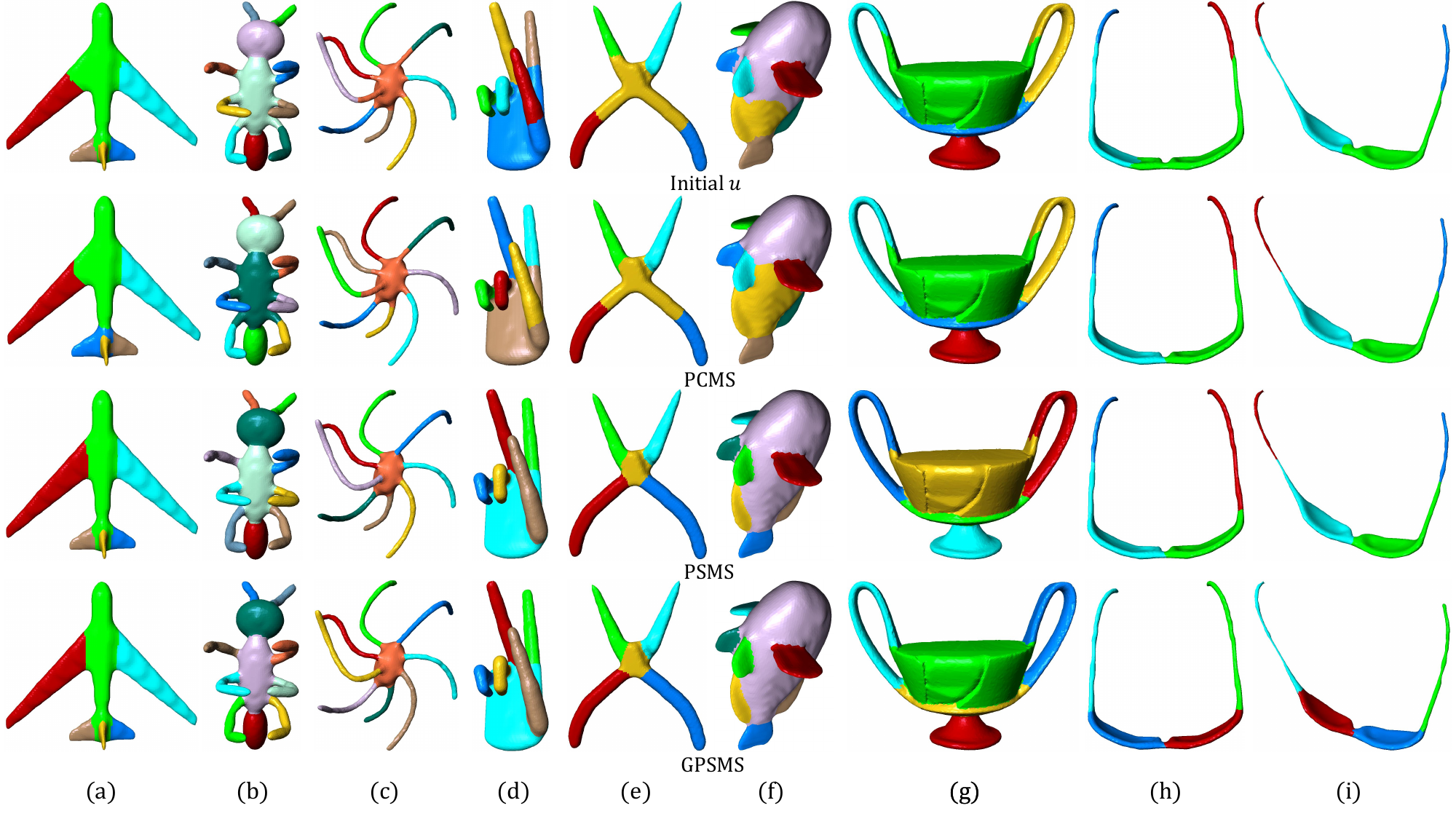}
\caption{{
Comparisons of multi-region segmentation results. $\mathrm{Initialization}$ (The first row), $\mathrm{PCMS}$ (The second row), $\mathrm{PSMS}$ (The third row) and $\mathrm{GPSMS}$ (The last row).
}
}\label{compare_PCMS}
\end{figure}

Specifically, with the same initializations, we present the comparisons of the multi-region segmentation results produced by the three methods in Fig.~\ref{compare_PCMS}. As observed, for the meshes in Fig.~\ref{compare_PCMS}, the segmentation boundaries of the PCMS method are not perceptually good. The PCMS method is sensitive to geometric structures, and  easily falls into local minima. It thus  failed  in getting the better boundaries (see the second row of Fig.~\ref{compare_PCMS}).  By introducing a new smooth function $\mathbf{b}$  to the PCMS method, the PSMS method is able to alleviate  the local minima case and find the better segmentation boundaries.  Most segmentation results of the PSMS method is better than that of the PCMS method (see (a-f) of the third row of Fig.~\ref{compare_PCMS}). In addition, as known,  both the PCMS method and the PSMS method use the total variation regularization, which aims
to get the shortest segmentation boundary. However, the shortest boundary is not always  the optimal boundary (for example the mesh (g),(h),(i) in Fig.~\ref{compare_PCMS}).  Therefore, for meshes with extreme irregular structures (for examples the meshes (g-i) in Fig.~\ref{compare_PCMS}),
the PCMS method and the PSMS method fail in describe the structures.  In contrast, the GPSMS method considers using a more general total variation regularization (the relaxed total generalized variation regularization), which involves second order differential operators and be able to capture more structures and overcome the staircase artifacts of TV regularization. It can relieve the influence of local minima and improve the segmentation boundaries.  As can bee seen from Fig.~\ref{compare_PCMS},  the segmentation results of the GPSMS method are all satisfying (see the last row of Fig.~\ref{compare_PCMS}). For the meshes in (a-f), the results of the PSMS method and the GPSMS method are similar, and are better than that of the PCMS method.  For the meshes in (g-i),  the segmentation results of the GPSMS method are better than that of the PCMS method and the PSMS method.
In all, the GPSMS method displays the ability to depict the different levels of  discontinuous information of the feature function of a mesh, and are  more suitable for dealing with complex meshes.

\subsection{Comparisons to other existing methods}

In the section, we tested our GPSMS segmentation results on the Princeton Segmentation Benchmark \cite{Chen09}, which contains 19 categories of 380 meshes for evaluating segmentation methods.
With this benchmark, we also compare our approach both quantitatively and visually to  Isoline Cut \cite{Oscar12},  WCSeg \cite{Kaick14}, PCMS \cite{zhang2018new}, Spectral $L_0$ \cite{tong2018spectral}, SB19 \cite{Kalogerakis10} and PCN12 \cite{Kalogerakis17}. For evaluating the segmentation results,  four evaluation metrics in \cite{Chen09} are used.

\begin{figure}[htbp]
\centering\includegraphics*[width=4.5in]{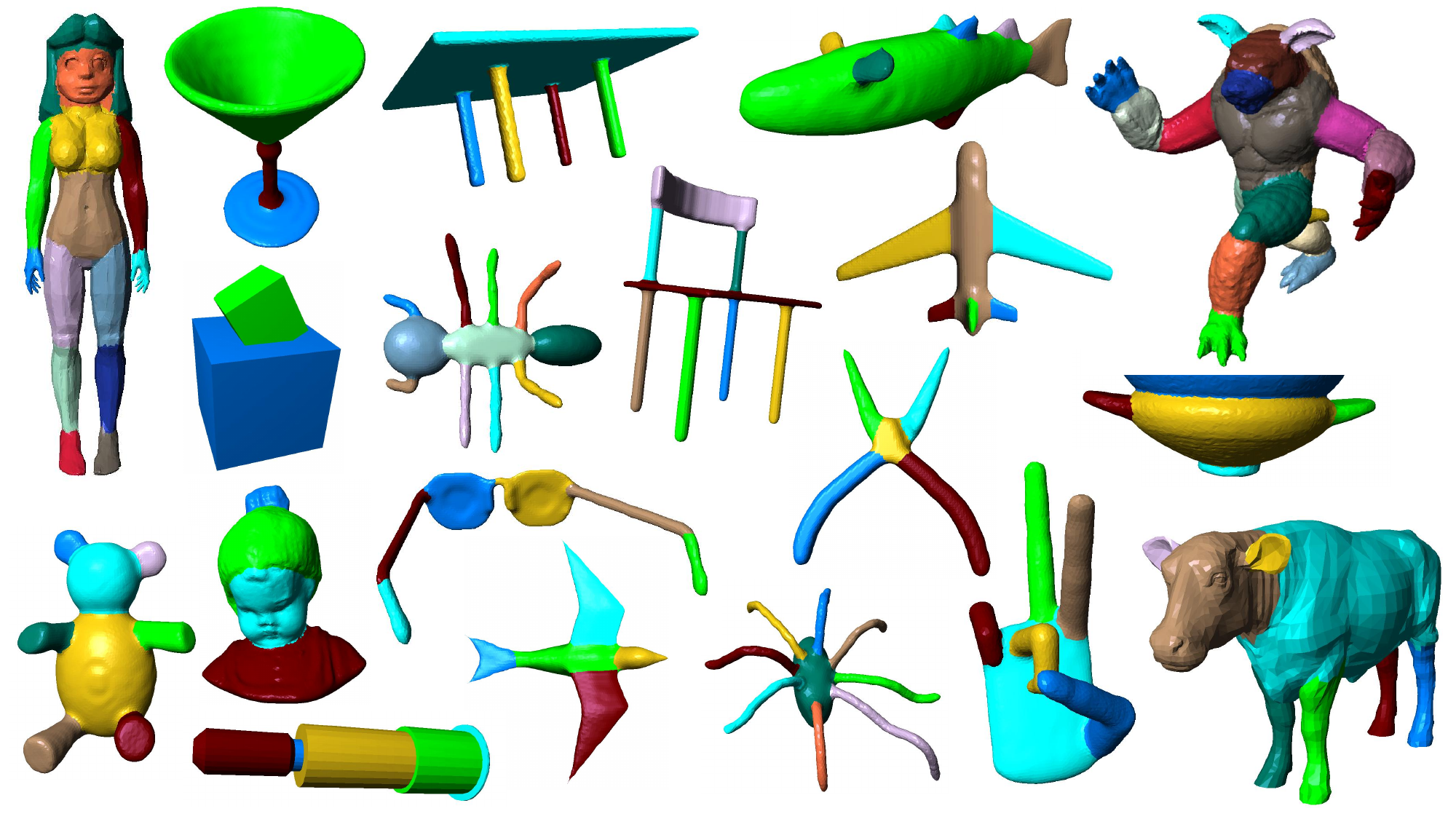}
\caption{{
Representative segmentation results of  19 categories in Princeton Segmentation Benchmark by our $\mathrm{GPSMS}$ method.
}
}\label{benchmark_result}
\end{figure}

Figure \ref{benchmark_result} shows the results by our GPSMS method for representative meshes of the 19 categories in the Benchmark. As can be seen, our results are consistent with the semantics of human perception.

\begin{figure}[htbp]
\centering\includegraphics*[width=4.5in]{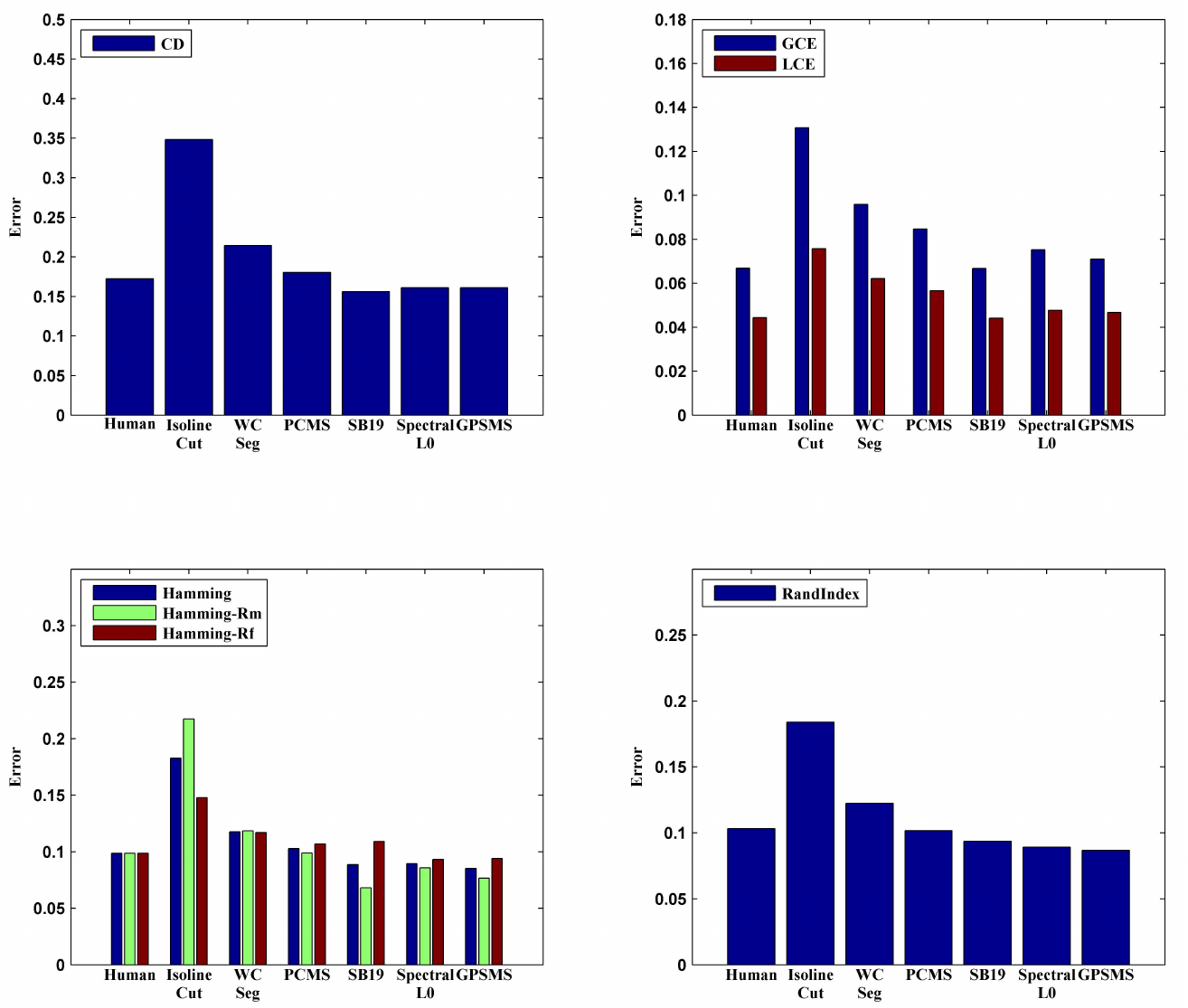}
\caption{
Evaluation of segmentation in terms of four evaluation metrics. The evaluations are performed  according to the protocols and human segmentations
in the Princeton Segmentation Benchmark of \protect\cite{Chen09}.
}\label{score}
\end{figure}

\begin{figure}[htbp]
\centering\includegraphics*[width=4.5in]{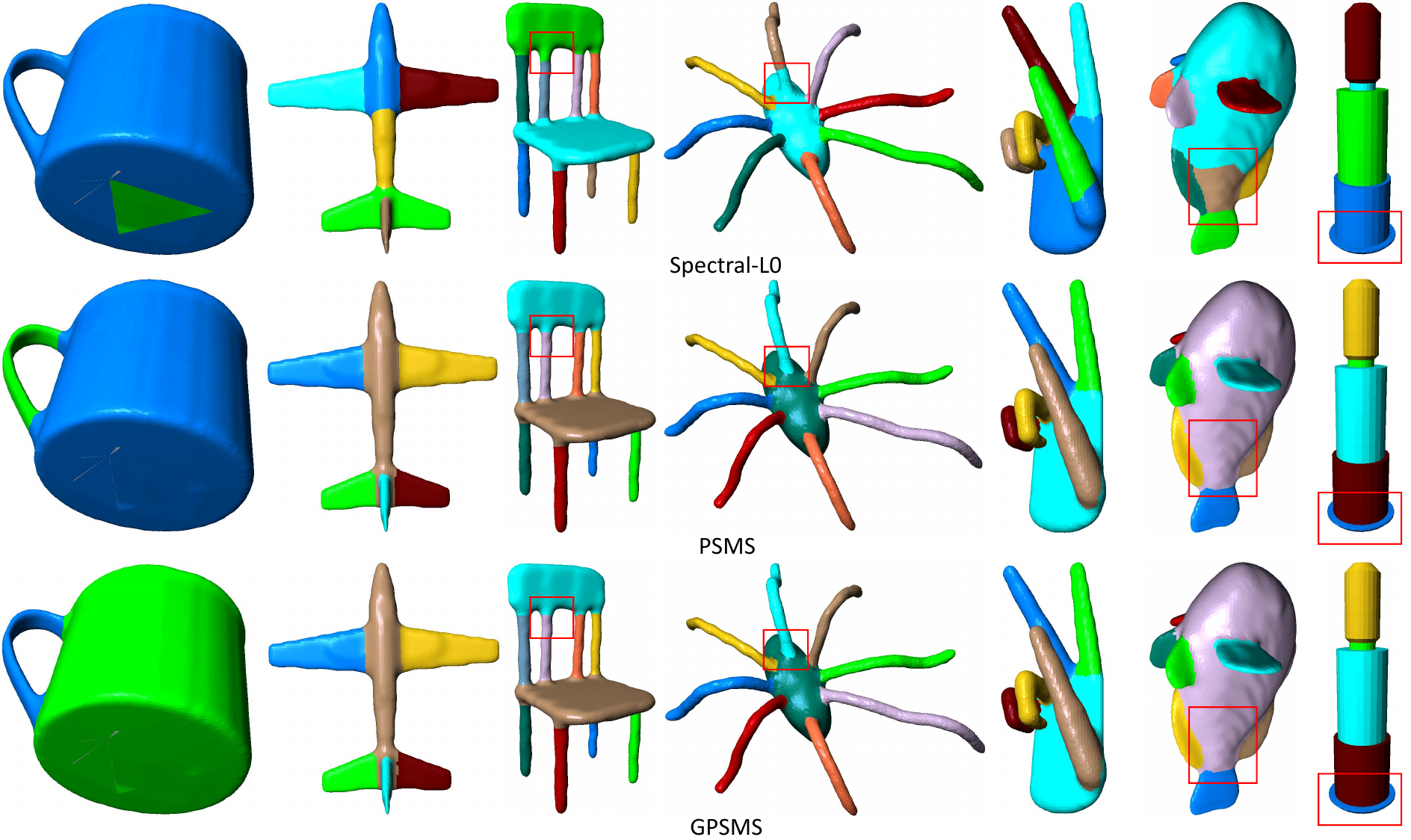}
\caption{
Comparison results between our segmentation method and the Spectral $L_0$ method \cite{tong2018spectral}.
}\label{compare_spectralL0}
\end{figure}

Fig.~\ref{score} shows the quantitative comparisons between our GPSMS method and other methods according to the four error metrics in \cite{Chen09}.  More details on the Rand Index (RI) metric are shown in Table~\ref{RandIndex}, where the RI scores of all the compared methods for the 19 categories of meshes in Princeton Segmentation Benchmark are listed. We can see that the average RI errors of   our results are lower  than those of other methods.
Fig.~\ref{compare_spectralL0} further present the comparisons of segmentation results between our method and  the Spectral $L_0$ method, respectively. As can be seen, the Spectral $L_0$ method either over segments the meshes or fails to capture the better boundaries (see the images marked by the red frames).  By contrast, our method can  produce results with better semantic parts and segmentation boundaries.
\begin{table*}[htpb]
\caption{Rand Index scores(the smaller the value, the better the segmentation results) across all 19 categories using our method and previous methods based on the Princeton Segmentation Benchmark of \protect\cite{Chen09}. }\label{RandIndex}
\centering
\begin{tabular}{lcccccccc}
\hline\noalign{\smallskip}
\multirow{1}{1.0cm}{Categories}  &  \multirow{1}{1.0cm}{Human}  & \multirow{1}{1.3cm}{IsolineCut}  & \multirow{1}{1.0cm}{WCSeg}  &  \multirow{1}{1.0cm}{PCMS} &\multirow{1}{1.0cm}{SB19}  &  \multirow{1}{1.3cm}{SpectralL0}  &\multirow{1}{1.0cm}{GPSMS} \\
\noalign{\smallskip}\hline\noalign{\smallskip}

 Human            & 13.8   & 12.5   &12.8 &13.0  &\textbf{12.0}  &12.84  &12.83 \\
   Cup            & 13.6   & 26.1   &17.1 &10.1  &10.0  &11.05   &\textbf{9.37}  \\
Glasses           & 10.1   & 13.3  &17.3 &9.4  &13.6 &\textbf{7.0}  &8.07    \\
 Airplane         & 9.2    & 19.7  &8.9  &10.5  &\textbf{7.9}  &8.4   &8.39 \\
 Ant              & 3.0    &4.5    &2.1   &2.2   &\textbf{1.8}   &1.86    &1.87  \\
Chair             & 8.9    &9.0   &10.3  &6.8   &\textbf{5.4}  &5.68   &6.19  \\
Octopus           & 2.4    & 6.4   &2.9   &2.9   &\textbf{1.9}    &2.18   &2.00   \\
Table             & 9.3    & 13.5  &9.1  &6.6   &\textbf{5.6}  &6.0   &6.04   \\
Teddy             & 4.9    & 10.1   & 5.6  &3.6  &\textbf{3.2}  &3.65   &3.57   \\
Hand              & 9.1    & 17.4   &11.6  &7.9  &10.5  &7.5   &\textbf{7.36}  \\
Plier             & 7.1    & 19.5  &8.7  &8.4  &\textbf{5.3}   &5.49   &5.98   \\
Fish              & 15.5   & 30.2  &20.3  &21.7  &\textbf{12.8}  &16.1   &13.23  \\
Bird              & 6.2   & 19.5   & 9.6 &7.5  &8.8  &\textbf{5.7}   &6.89    \\
Armadillo         & 8.3    & 17.9   &8.1   &9.2  &\textbf{7.0}  &7.6    &8.17   \\
Bust              & 22.0   & 34.5  &26.6  &24.8  &23.8 &22.0   &\textbf{21.7}  \\
Mech              & 13.1   & 28.4  &18.2  &9.2  &11.0  &9.0   &\textbf{8.96}  \\
Bearing           & 10.4   & 28.6  &11.9  &10.7  &9.1  &8.6   &\textbf{8.06}  \\
Vase              & 14.4   & 18.0  &16.1  &12.6  &14.7  &\textbf{10.8}   &11.18  \\
FourLeg           & 14.9   & 20.3  &15.2  &15.9  &\textbf{13.8}  &17.76   &14.83  \\
\textbf{Average } & 10.3   & 18.4  &12.2  &10.2  &9.4  &8.91   &\textbf{8.67}  \\
\noalign{\smallskip}\hline
\end{tabular}
\end{table*}

\begin{figure}[htbp]
\centering\includegraphics*[width=4.5in]{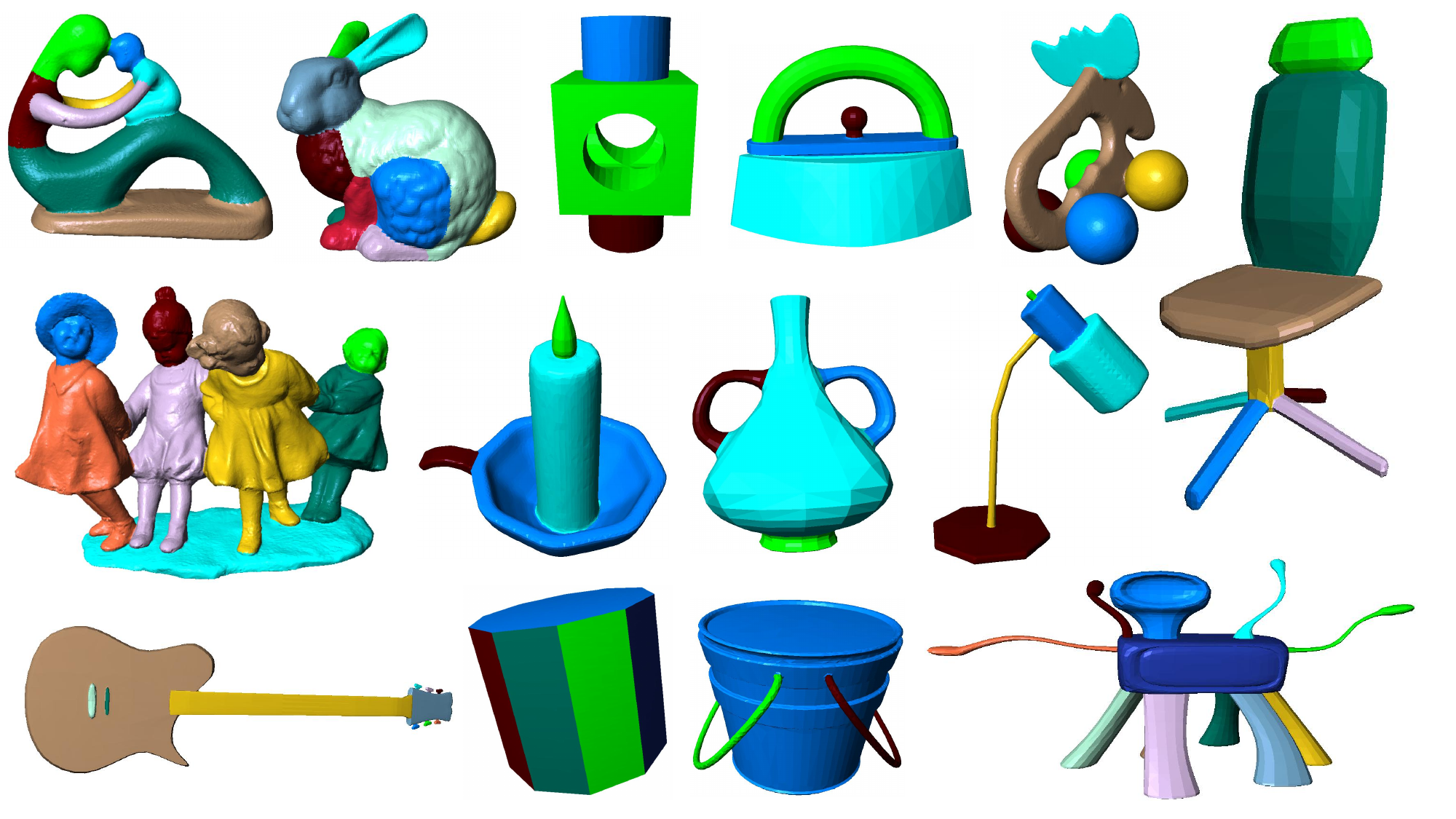}
\caption{{
The segmentation results of other models  by our $\mathrm{GPSMS}$ method .
}
}\label{complex_result}
\end{figure}

In addition to the benchmark data,  we tested our algorithm on more meshes in Fig.~\ref{complex_result}, and our method can get satisfying results.
\begin{table}[htbp]
\caption{Comparison of computational  time (seconds) between  our segmentation method and the $\mathrm{PCMS}$ method for surfaces in Fig.~\ref{compare_PCMS}}\label{time_table}
\centering
\begin{tabular}{|l|l|l|l|l|l|}
\hline
\multirow{2}{*}{Mesh}  & \multirow{2}{*}{$\mathrm{T}$}  & \multirow{2}{*}{\#Parts}  &  \multirow{2}{*}{PCMS} &  \multirow{2}{*}{PSMS} &  \multirow{2}{*}{GPSMS} \\
& &  & & &\\
\hline
  Fig.~\ref{compare_PCMS} (a)          & 11162  & 6    & 0.24  & 0.83 & 1.9    \\
  \hline
  Fig.~\ref{compare_PCMS} (b)          &10624   & 11  &1.4   &2.3  & 7.1 \\
 \hline
 Fig.~\ref{compare_PCMS}  (c)           & 12646  & 9   & 0.21   &1.1  & 2.4 \\
\hline
 Fig.~\ref{compare_PCMS} (d)           & 17044 &6  & 0.4  & 1.5 &  4.6    \\
\hline
\end{tabular}
\end{table}
\begin{figure}[htbp]
\centering\includegraphics*[width=4.5in]{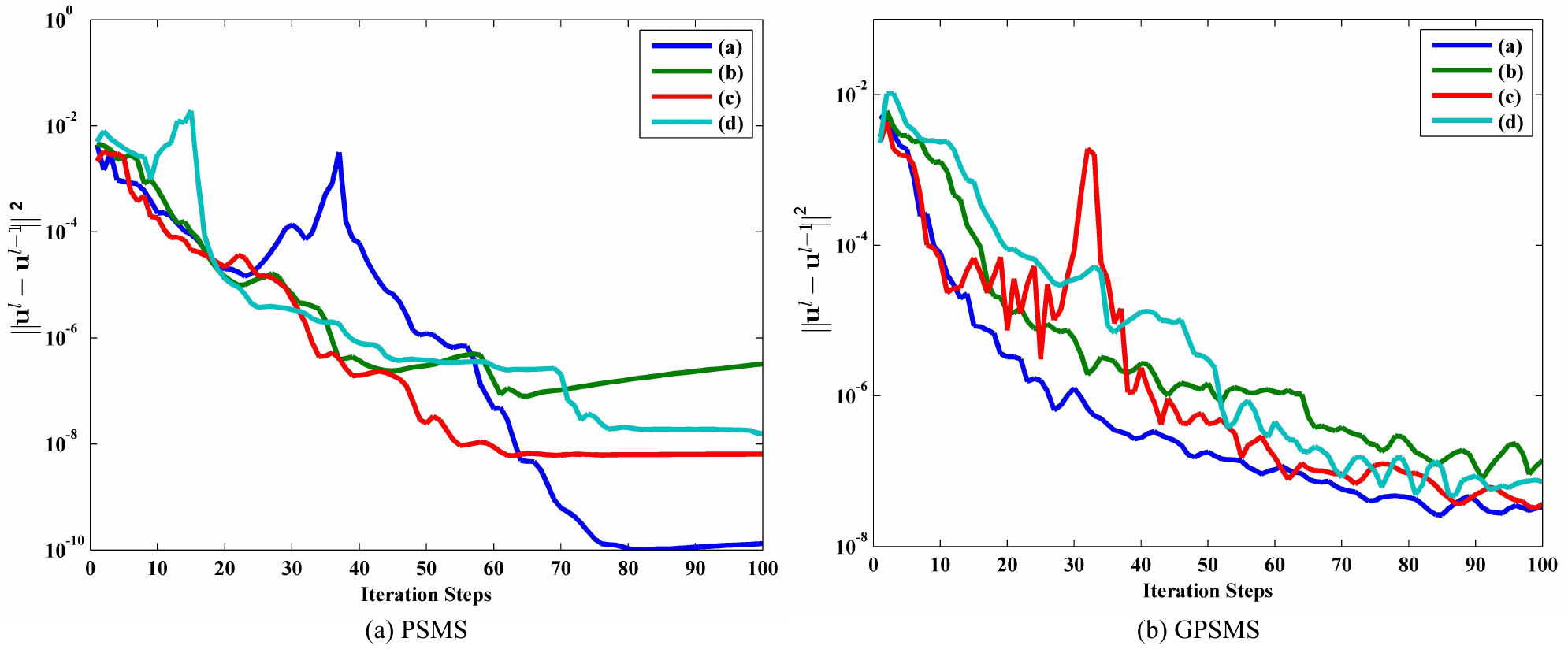}
\caption{{
The absolute error $\|\mathbf{u}^{n+1}-\mathbf{u}^{n}\|^2_{\mathbf{U}_{M}}$ with 100 iterations for examples (a-d) in Fig.~\ref{compare_PCMS}  by our $\mathrm{PSMS}$ method and $\mathrm{GPSMS}$ method, respectively.
}
}\label{error_compare}
\end{figure}

\subsection{Computational cost}
We now discuss the computational cost. The main  computational cost is the iteratively  solving the six sub problems in Algorithm~\ref{algorithm2}. The main factors affecting the computational cost are mesh size, segmentation number and the parameter $\alpha$. Meshes with large-size meshes or more segmentation parts consume longer computational time. We also find that if $\alpha$ is too small, the algorithm will be slow.  Table~\ref{time_table} compares  the CPU costs between our methods (PSMS and GPSMS) and the PCMS method for examples  (a-d) in Fig.~\ref{compare_PCMS}. As our methods introduce the new smooth variable $\mathbf{b}$ and the RTGV regularization for the GPSMS method, the overall running time is longer than that of the PCMS method, especially for the GPSMS method.  In addition, Fig.~\ref{error_compare} presents the error curves with 100 iterations according to the absolute error $\|\mathbf{u}^l-\mathbf{u}^{l-1}\|^2_{\mathbf{U}_M}$ for results (a-d) in Fig.~\ref{compare_PCMS} by our PSMS method and GPSMS method, respectively. The error curves show that  the errors within 100 iterations decreases fast,  which also indicates that our algorithms are numerically robust.

\subsection{Limitations}

\begin{figure}[htbp]
\centering\includegraphics*[width=4.3in]{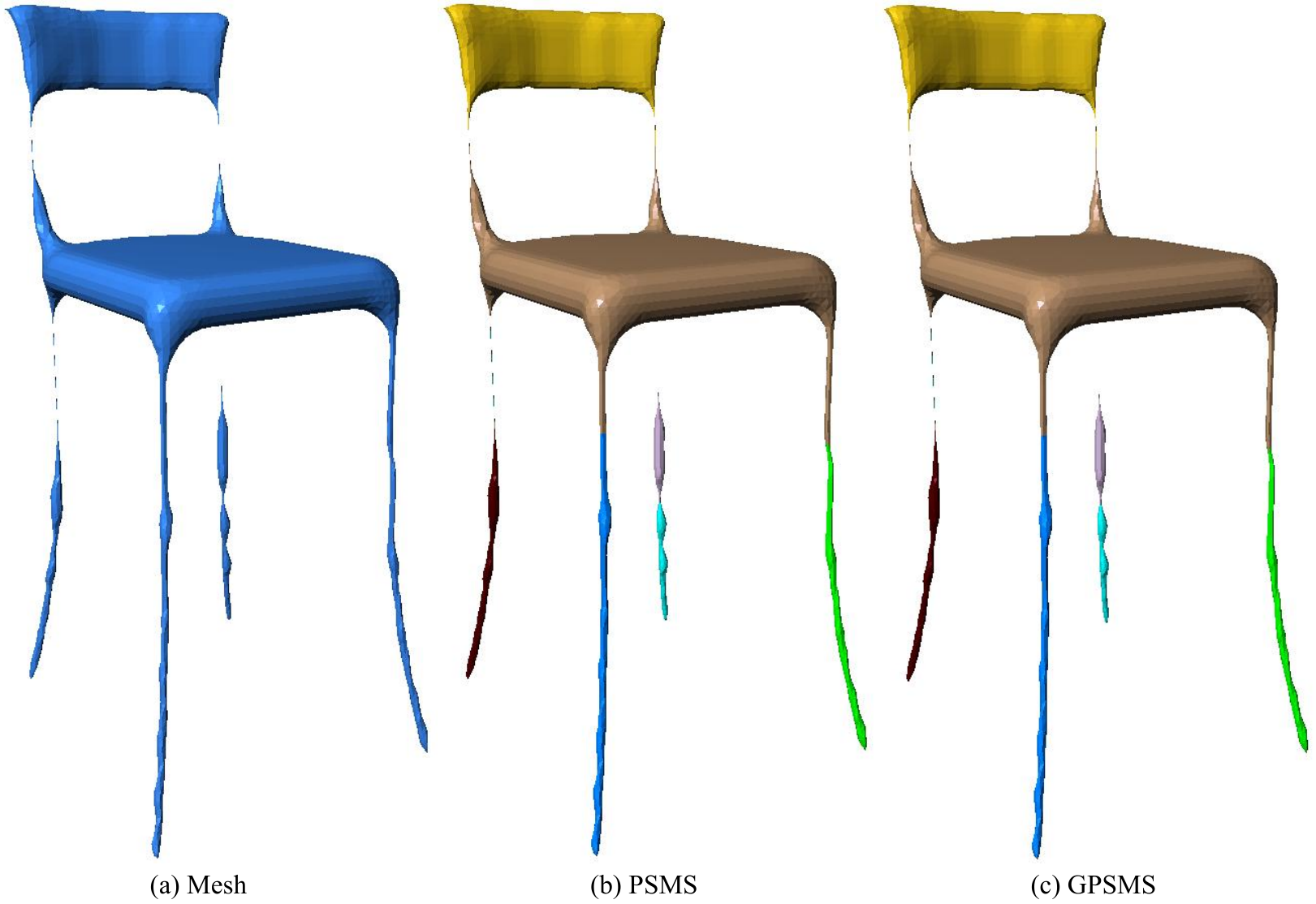}
\caption{{
  Failed results produced by  our  $\mathrm{PSMS}$ method and our $\mathrm{GPSMS}$ method.
}
}\label{limitation}
\end{figure}

Our method has been demonstrated very effective in getting the better segmentation boundaries, but it still has some limitations. Firstly, we cannot give a formula to precisely compute the parameters and segmentation parts $\mathrm{K}$. Secondly, for meshes with quite poor quality, our algorithm cannot give satisfying segmentation results; see Fig.~\ref{limitation} for an example. The mesh in Fig.~\ref{limitation}(a) has very poor quality, in this case, we cannot get correct results (see Fig.~\ref{limitation}(b-c)). Finally, as  the Mumford-Shah model is non-convex,  our method is thus affected by the clustering center $\mu$ due to the variant of the MS model.

\section{\label{conclusion}Conclusions}

The Mumford-Shah (MS) model is a vital tool for mesh segmentation, which pursues the shortest length of boundaries by the total variation regularization. The existing efforts try to solve the MS model through approximating the solutions with the  piecewise constant functions. Different from  most previous methods, in the article, we presented a novel piecewise smooth Mumford-Shah  mesh segmentation technique using the total generalized variation regularization. The  smooth model assumes to approximate the solutions by a sum of piecewise constant functions and a smooth function, and  the TGV regularization involves the high order differential operator to
  characterize the high order discontinuous structures of the mesh. The proposed mesh segmentation method is effective in segmenting meshes with irregular structures and getting the better boundaries.
 We solve our piecewise smooth MS method by the optimization  iterative algorithm  based on alternating minimization and ADMM, where the subproblems are solved  by either   the closed form solution or linear systems.  Our algorithm is discussed from several aspects, including setting of parameters, computational costs, comparisons with piecewise constant MS model and evaluation on the Princeton Segmentation Benchmark with several state-of-art methods. Experimental results show that our piecewise smooth MS method can yield competitive results when compared to other approaches.  Furthermore, the quantitative errors and computational costs further confirm the robustness and efficiency of the proposed method.

There are a few problems for further investigation. For instance,  the preprocessing of  the feature function of the mesh by the manifold learning techniques and the point cloud processing by the piecewise smooth MS model.

\section*{Acknowledgments}
We would like to thank Pengfei Xu, Noa Fish, Weihua Tong for providing
their segmentation data of \cite{Oscar12}, \cite{Kaick14} , \cite{tong2018spectral}, and the Princeton Segmentation Benchmark.
The work is supported by the fund of Beijing Polytechnic,2023X005-KXD.

\bibliography{egbibsample}

%

\appendix
\subsection*{Proof of Theorem 1}
\begin{proof}
As  the energy $E$ is positive, proper, and bounded from below by zero, let $(\mathbf{u}^n,\mathbf{b}^n,\mu^n)$ be a minimizing sequence of \eqref{ps_MS_Mesh} that satisfies $$E(\mathbf{u}^n,\mathbf{b}^n,\mu^n) \rightarrow \mathrm{inf} E(\mathbf{u,b},\mu),\  n\rightarrow\infty.$$
Then, there exists a  $M > 0$ such that  $E(\mathbf{u}^n,\mathbf{b}^n,\mu^n) \leq M$.   This implies
that each term in $E(\mathbf{u}^n,\mathbf{b}^n,\mu^n)$ is bounded. That is
$$ \mathrm{GTV}^p(\mathbf{u}) \leq M, \ \  \|\triangle\mathbf{b}\|^2 \leq M, \|\mathbf{b}\|^2 \leq M, \  \langle\mathbf{u},s(\mathbf{f},\mathbf{b},\mu)\rangle_{\mathbf{U}_M} \leq M. $$
Moreover, since  $\mathbf{u}^n_k \in[0,1]$ , we conclude that $\{\mathbf{u}^n\}$ is bounded in $\mathrm{BV}(\Omega)$. By the compact embedding theorem, there exists a  subsequence $\{\mathbf{u}^{n_{l}}\}$  convergent to $\mathbf{u}^\ast$ in $L^1(\Omega)$.

Meanwhile, as $\{\mathbf{\mu}_k^n\}$ is bounded, we can get a subsequence $\{\mathbf{\mu}_k^{n_l}\}$ converge to a constant $\{\mathbf{\mu}_k^\ast\}$.

In addition,  we note that
  $$\int_\Omega\mid\mathbf{b}^n\mid^2 dx < M, \ \ \  \int_\Omega \mid\triangle\mathbf{b}^n\mid^2 dx < M.$$
  This show that $\{\mathbf{b}^n\}$  is also bounded. Therefore, it exists a subsequence $\{\mathbf{b}^{n_l}\}$ that converges to $\mathbf{b}^\ast $ in $L^2(\Omega)$.
  Then, by the lower semi-continuity, we have
  $$\mathrm{GTV}^p(\mathbf{u}^\ast) \leq \liminf\limits_{n\rightarrow\infty} \mathrm{GTV}^p(\mathbf{u}^n),$$
  $$\|\triangle\mathbf{b}^\ast\|^2 \leq \liminf\limits_{n\rightarrow\infty} \|\triangle\mathbf{b}^n\|^2,$$
   $$\|\mathbf{b}^\ast\|^2 \leq \liminf\limits_{n\rightarrow\infty} \|\mathbf{b}^n\|^2.$$
 Moreover, by the Fatou's lemma,  we get
  $$\langle\mathbf{u}^\ast,s(\mathbf{f},\mathbf{b}^\ast,\mu^\ast)\rangle_{\mathbf{U}_M}\leq \liminf\limits_{n\rightarrow\infty} \leq \langle\mathbf{u}^n,s(\mathbf{f},\mathbf{b}^n,\mu^n)\rangle_{\mathbf{U}_M}.$$
  We then have
  $$E(\mathbf{u}^\ast,\mathbf{b}^\ast,\mathbf{\mu}^\ast)\leq \liminf\limits_{n\rightarrow\infty}E(\mathbf{u}^n,\mathbf{b}^n,\mu^n) = \inf E(\mathbf{u,b,\mu}),$$
  which implies that $(\mathbf{u}^\ast,\mathbf{b}^\ast,\mathbf{\mu}^\ast)$ is a minimizer of problem \eqref{ps_MS_Mesh}.
\end{proof}

\subsection*{Proof of Theorem 2}
\begin{proof}
 By the Lagrange multiplier formulas, we have
 \begin{equation}
 \begin{split}
  &\lim_{l\rightarrow\infty} r_\mathbf{p}^{-1}(\lambda_\mathbf{p}^{l}-\lambda_\mathbf{p}^{l-1}) = \lim_{k\rightarrow\infty} \mathbf{p}^l-(\nabla \mathbf{u}^l-\mathbf{v}^l)
 =\mathbf{p}^\ast-(\nabla \mathbf{u}^\ast-\mathbf{v}^\ast)=0,\\
 &\lim_{l\rightarrow\infty} r_\mathbf{q}^{-1}(\lambda_\mathbf{q}^l-\lambda_\mathbf{q}^{l-1}) = \lim_{k\rightarrow\infty} \mathbf{q}^l-\mathrm{div}(w\mathbf{v}^l)
 =\mathbf{q}^\ast-\mathrm{div}(w\mathbf{v}^\ast)=0,\\
 &\lim_{l\rightarrow\infty} r_\mathbf{z}^{-1}(\lambda_\mathbf{z}^l-\lambda_\mathbf{z}^{l-1}) = \lim_{k\rightarrow\infty} \mathbf{z}^l-\mathbf{u}^l
 =\mathbf{z}^\ast-\mathbf{u}^\ast=0.
 \end{split}
 \end{equation}
The optimization conditions of $\mathbf{u,v,b,\mu}$ sub-problems are
 \begin{equation}
 \begin{split}
  &-\lambda_\mathbf{z}^{l-1} +r_\mathbf{z} (\mathbf{u}^l-\mathbf{z}^{l-1}) + \mathrm{div}(\lambda_\mathbf{p}^{l-1}) -r_\mathbf{p}\mathrm{div}(\nabla \mathbf{u}^l-\mathbf{v}^{l-1}-\mathbf{p}^{l-1})=0,\\
  &\lambda_\mathbf{p}^{l-1} + r_\mathbf{p}(\mathbf{v}^l-(\nabla \mathbf{u}^l-\mathbf{p}^{l-1})) +\nabla(\lambda_\mathbf{q}^{l-1}) - r_\mathbf{q} \nabla(\mathrm{div}(\mathbf{v}^l)-\mathbf{q}^{l-1}) = 0,\\
  &(\beta\Delta^2+\eta\mathcal{I} + \alpha\mathcal{I})\mathbf{b}^l +  \alpha<-(\mathbf{f}-\mu^{l-1}),\mathbf{z}^{l-1}> = 0,\\
  &\alpha<\mu^l-(\mathbf{f}-\mathbf{b}^{l}),\mathbf{u}^{l}> = 0.
 \end{split}
 \end{equation}
 We rearrange above formulas and get
  \begin{equation}
    \begin{split}
  &-\lambda_\mathbf{z}^{l-1} +r_\mathbf{z} (\mathbf{u}^l-\mathbf{u}^{l-1} + \mathbf{u}^{l-1}-\mathbf{z}^{l-1}) + \mathrm{div}(\lambda_\mathbf{p}^{l-1}) \\
  &\ \ \ \ \ \ \ \  \ \ \ \ -r_\mathbf{p}\mathrm{div}(\nabla \mathbf{u}^l-\mathbf{v}^{l}-\mathbf{p}^{l}+\mathbf{v}^{l}-\mathbf{v}^{l-1}+\mathbf{p}^{l}-\mathbf{p}^{l-1})=0,\\
  &\lambda_\mathbf{p}^{l-1} + r_\mathbf{p}(\mathbf{v}^l-(\nabla \mathbf{u}^l-\mathbf{p}^{l} +\mathbf{p}^{l}-\mathbf{p}^{l-1})) + \nabla(\lambda_\mathbf{q}^{l-1}) \\
  &\ \ \ \ \ \ \ \ \ \ \ \ \ - r_\mathbf{q} \nabla(\mathrm{div}(\mathbf{v}^l)-\mathbf{q}^{l}+\mathbf{q}^{l}-\mathbf{q}^{l-1}) = 0,\\
  &(\beta\Delta^2+\eta\mathcal{I} + \alpha\mathcal{I})\mathbf{b}^l + \alpha<-(\mathbf{f}-\mu^{l}+\mu^{l}-\mu^{l-1}),\mathbf{z}^{l}-\mathbf{z}^{l}+\mathbf{z}^{l-1}> = 0,\\
  & \alpha<\mu^l-(\mathbf{f}-\mathbf{b}^{l}),\mathbf{u}^{l}> = 0.
  \end{split}
 \end{equation}
By taking the limit $l\rightarrow\infty$, we get
 \begin{equation}
    \begin{split}
     &-\lambda_\mathbf{z}^\ast+ \mathrm{div}(\lambda_\mathbf{p}^\ast)=0,  \lambda_\mathbf{p}^\ast +w\nabla(\lambda_\mathbf{q}^\ast)  = 0,\\
  &(\beta\Delta^2+\eta\mathcal{I} + \alpha\mathcal{I})\mathbf{b}^\ast +  \alpha<-(\mathbf{f}-\mu^\ast),\mathbf{z}^\ast> = 0,\\
  & \alpha<\mu^\ast-(\mathbf{f}-\mathbf{b}^\ast),\mathbf{u}^\ast> = 0.
   \end{split}
 \end{equation}
 Next, we consider the optimality conditions of $\mathbf{p,q,z}$ sub-problems
\begin{equation}
 \begin{split}
 &\partial(\|\mathbf{p}^l\|_1)+ \lambda_\mathbf{p}^{l-1} + r_\mathbf{p}(\mathbf{p}^l - (\nabla\mathbf{u}^l - \mathbf{v}^l)) \ni 0,\\
 &\alpha_0\partial(\|\mathbf{q}^l\|_1)+ \lambda_\mathbf{q}^{l-1} + r_\mathbf{q}(\mathbf{q}^l - \mathrm{div}(\mathbf{v}^l) \ni 0,\\
 &\alpha s(\mathbf{f},\mathbf{b}^l,\mu^l) +\partial\mathcal{X}(\mathbf{z}^l) + \lambda_\mathbf{z}^l + r_\mathbf{z}(\mathbf{z}^l - \mathbf{u}^l) \ni 0.
 \end{split}
 \end{equation}

 Then, in the limits $k\rightarrow\infty$, we obtain
 \begin{equation}
 \begin{split}
 &\partial(\|\mathbf{p}^\ast\|_1)+ \lambda_\mathbf{p}^\ast \ni 0,\\
 &\alpha_0\partial(\|\mathbf{q}^\ast\|_1)+ \lambda_\mathbf{q}^\ast \ni 0,\\
 &\alpha s(\mathbf{f},\mathbf{b}^\ast,\mu^\ast) +\partial\mathcal{X}(\mathbf{z}^\ast) + \lambda_\mathbf{z}^\ast \ni 0.
 \end{split}
 \end{equation}
\end{proof}

\end{document}